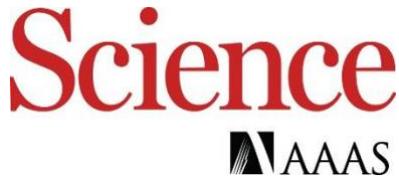

# Title: CEERS: Forging the First Dust – Transition from Stellar to ISM Grain Growth in the Early Universe

Authors: Denis Burgarella[1*], Véronique Buat[1], Patrice Theulé[1], Jorge Zavala[2], Pablo Arrabal Haro[3], Micaela B. Bagley[4], Médéric Boquien[5], Nikko Cleri[6,7,8], Tim Dewachter[1], Mark Dickinson[3], Henry C. Ferguson[9], Vital Fernández[10], Steven L. Finkelstein[11], Adriano Fontana[12], Eric Gawiser[13], Andrea Grazian[14], Norman Grogin[9], Benne W. Holwerda[15], Jeyhan S. Kartaltepe[16], Lisa Kewley[17], Allison Kirkpatrick[18], Dale Kocevski[19], Anton M. Koekemoer[9], Arianna Long[4], Jennifer Lotz[9], Ray A. Lucas[9], Bahram Mobasher[20], Casey Papovich[21,22], Pablo G. Pérez-González[23], Nor Pirzkal[24], Swara Ravindranath[25,26], Giulia Rodighiero[27,28], Yannick Roehlly[1], Caitlin Rose[16], Lise-Marie Seillé[1], Rachel Somerville[29], Steve Wilkins[30,31], Guang Yang[32], L. Y. Aaron Yung[9]

**Affiliations:**

[1]Aix Marseille Université, CNRS, CNES, LAM, Marseille, France

[2]National Astronomical Observatory of Japan, 2-21-1 Osawa, Mitaka, Tokyo 181-8588, Japan

[3]NSF's National Optical-Infrared Astronomy Research Laboratory, 950 N. Cherry Ave., Tucson, AZ 85719, USA

[4]Department of Astronomy, The University of Texas at Austin, Austin, TX, USA

[5]Université Côte d'Azur, Observatoire de la Côte d'Azur, CNRS, Laboratoire Lagrange, 06000, Nice, France

[6]Department of Astronomy and Astrophysics, The Pennsylvania State University, University Park, PA 16802, USA

[7]Institute for Computational and Data Sciences, The Pennsylvania State University, University Park, PA 16802, USA

[8]Institute for Gravitation and the Cosmos, The Pennsylvania State University, University Park, PA 16802, USA

[9]Space Telescope Science Institute, Baltimore, MD, USA

[10]Instituto de Investigación Multidisciplinar en Ciencia y Tecnología, Universidad de La Serena, Raul Bitràn 1305, La Serena 2204000, Chile

[11]Department of Astronomy, The University of Texas at Austin, Austin, TX, USA

[12]INAF - Osservatorio Astronomico di Roma, via di Frascati 33, 00078 Monte Porzio Catone, Italy

[13]Department of Physics and Astronomy, Rutgers, the State University of New Jersey, Piscataway, NJ 08854, USA

[14]INAF--Osservatorio Astronomico di Padova, Vicolo dell'Osservatorio 5, I-35122, Padova, Italy

[15] Physics & Astronomy Department, University of Louisville, 40292 KY, Louisville, USA

[16] Laboratory for Multiwavelength Astrophysics, School of Physics and Astronomy, Rochester Institute of Technology, 84 Lomb Memorial Drive, Rochester, NY 14623, USA

[17] Center for Astrophysics | Harvard & Smithsonian, 60 Garden Street, Cambridge, MA 02138, USA

[18] Department of Physics and Astronomy, University of Kansas, Lawrence, KS 66045, USA

[19] Department of Physics and Astronomy, Colby College, Waterville, ME 04901, USA

[20] Department of Physics and Astronomy, University of California, 900 University Ave, Riverside, CA 92521, USA

[21] Department of Physics and Astronomy, Texas A&M University, College Station, TX, 77843-4242 USA

[22] George P. and Cynthia Woods Mitchell Institute for Fundamental Physics and Astronomy, Texas A&M University, College Station, TX, 77843-4242 USA

[23] Centro de Astrobiologia (CAB), CSIC-INTA, Ctra. de Ajalvir km 4, Torrejon de Ardoz, E-28850, Madrid, Spain

[24] ESA/AURA Space Telescope Science Institute

[25] Astrophysics Science Division, NASA Goddard Space Flight Center, 8800 Greenbelt Road, Greenbelt, MD 20771, USA

[26] Center for Research and Exploration in Space Science and Technology II, Department of Physics, Catholic University of America, 620 Michigan Ave N.E., Washington DC 20064, USA

[27] Department of Physics and Astronomy, Università degli Studi di Padova, Vicolo dell'Osservatorio 3, I-35122, Padova, Italy

[28] INAF - Osservatorio Astronomico di Padova, Vicolo dell'Osservatorio 5, I-35122, Padova

[29] Center for Computational Astrophysics, Flatiron Institute, 162 5th Avenue, New York, NY, 10010, USA

[30] Astronomy Centre, University of Sussex, Falmer, Brighton BN1 9QH, UK

[31] Institute of Space Sciences and Astronomy, University of Malta, Msida MSD 2080, Malta

[32] Nanjing Institute of Astronomical Optics and Technology, Nanjing 210042, China

*Corresponding author. Email: denis.burgarella@lam.fr

**Abstract:** We investigate the coevolution of metals and dust for 173 galaxies at $4.0<z<11.4$ observed with JWST/NIRSpec. We use the code CIGALE that integrates photometric and spectroscopic data. Our analysis reveals a critical transition at $M_{star}\sim10^{8.5}$ M$\odot$, from galaxies dominated by supernovae and AGB stardust, to those dominated by grain growth. This implies a two-mode building of dust mass, supported by model predictions. The detection of stardust galaxies provides a natural and inherent explanation to the excess of UV-bright galaxies at $z>10$ by JWST. Besides, we observe that the metallicity of galaxies at $z\gtrsim8$ presents a metal-to-stellar mass ratio larger than a few $10^{-3}$, above a floor. This suggests a very fast rise of metals at high redshift, impacting the tentative detections of population III objects.

**Main Text:**

Nucleosynthesis started in the first stellar population formed in the Universe: population III stars (pop.III), and shortly after pop.II stars. Then, supernovae (SNe) and asymptotic giant branch (AGB) stars expelled the first metals that formed the first dust grains (*1*, *2*, *3*, *4*). Even though dust makes up about 1% of the total interstellar medium (ISM) mass in galaxies (*5*), it is a fundamental component of their ISM. First, it impacts the ultraviolet (UV) and optical emissions through dust attenuation and reddening. This energy warms dust and is reemitted at infrared (IR) and sub-millimeter (submm) wavelengths (*6*, *7*). More crucial is the key role played by dust when cooling the ISM to form low-mass stars. Because the $H_2$ molecule, that regulates the cloud collapse and star formation, does not efficiently form in the gas phase under typical ISM conditions, the catalysis on the surface of the grains is fundamental to make this process efficient (*8*). Those grains are then used as seeds in the ISM for grain growth (*9*, *10*). Collisions between gas and dust grains efficiently cool the gas and lead to fragmentation at $n_H \sim 10^{12}$ cm$^{-3}$ and form low-mass stars, that is a transition between pop.III and pop.II (*11*).

JWST found an unpredicted excess of UV-luminous galaxies at z>10 compared to HST-calibrated models. The most popular explanations (*12*) for this excess are: a top-heavy initial mass function (IMF), or an origin related to dust: either a low dust attenuation (*13*), and/or a special dust/star morphology.

In the last decade, we have detected dusty galaxies at zε4 (*14*, *15*, *16*) with large dust masses (*5*, *17*) that cannot be explained by models. SNe and AGB stars scarcely could be at the origin of such a large dust mass, especially if we account for the reverse shock due to the expanding SN blast wave in the ISM (*18*). On the other hand, dust mass growth is regulated by a critical minimum ISM metallicity (*10*). The time to reach this critical metallicity depends on the star-formation timescale. It is about $10^8$<age[years]<$10^9$, that is generally longer than the age of the Universe for our sample of galaxies. Dust formation models (*19*, *20*, *21*) suggest a transition from galaxies only containing stardust created by SNe and AGB stars, to galaxies where grain growth by accretion of metals in interstellar clouds becomes dominant (*22*, *23*, *24*). However, this transition that should happen at relatively low stellar mass wad not observed, yet.

This paper derives new constraints on the ISM at high redshift ofrom the JWST/CEERS[1] project, featuring NIRCAM, NIRSpec, plus ancillary data from SCUBA-2 (*25*) and NOEMA (*15*).

1. **The NIRSpec prism spectroscopic sample**
   1.1. *The origin of the sample*

CEERS's NIRCam observations detect 101,808 objects photometrically (CEERS_v0.51.4, *26*). We also have 1,337 spectroscopic observations with NIRSpec (*27*, *28*). We use 634 of these NIRSpec observations carried out with the prism configuration. Whenever possible, we combine spectroscopic data (*29*) with photometric data by cross-matching the coordinates within 0.2 arcsec. However, some objects do not have any photometry but only spectroscopy. For them, we only use the spectroscopic data. After fitting the prism spectra (with and without NIRCam data), we check the quality of the spectroscopic redshifts for objects with $z_{spec}$>4.0. We classify the redshifts in four classes of redshift quality from $q_z=0$ (no doubts on redshift), to $q_z=4$ (wrong or unconfirmed redshift). We keep 173 objects with NIRSpec observations, for which $q_z=0$ (all modeled lines match the observed spectrum) or $q_z=1$ (some fainter lines not in excellent agreement with models). Among these objects, we identify a sample of 6 possible AGN (*30*, *31*, *32*). These AGN are listed in tab. S1, and flagged with crosses in the plots. The distribution of redshifts is shown in Fig. 1.

---

[1] The Cosmic Evolution Early Release Science Survey, https://ceers.github.io/

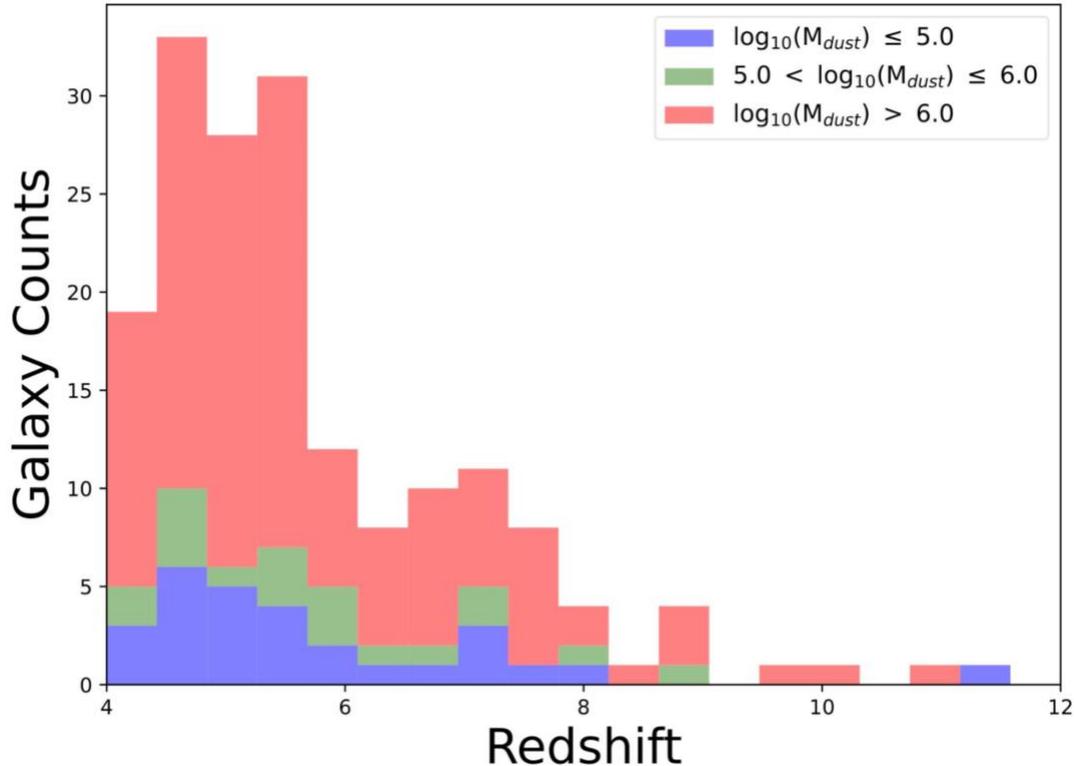

**Fig. 1: Distribution of spectroscopic redshifts.** derived by fitting the spectrophotometric data with CIGALE. We keep 173 objects with the most robust redshifts. The redshift distributions are presented for objects with $\log_{10}(M_{dust})\leq5.0$ (blue), $5.0<\log_{10}(M_{dust})\leq6.0$ (green), $\log_{10}(M_{dust})>6.0$ (red).

### *1.2. Analysis of the spectrophotometric data*

We build the spectral energy distributions (SEDs) using all photometric data with a signal-to-noise (SNR) ratio ε1.0. All the other measures are set as upper limits in the fit. We use a version of CIGALE that accepts both photometric and spectroscopic data (Supplementary Materials, Description of the spectro-photometric CIGALE). The priors used in the two fits are listed in tab. S2. A sample of the fits is shown (Supplementary Materials: Sample of spectral fits).

Two star-formation histories (SFHs) are used to test the stability of the results: a delayed-plus-burst[2], and a periodic[3] one. The delayed SFH assumes that star formation is active over a few tens to hundreds Myrs, with a final burst. Various other types of SFHs, including non-parametric ones (*34*, *35*, *36*, *37*) could be used, but determining the SFH in the early Universe is difficult for any SED modeling method (*38*). Moreover, (*34*, *35*, *36*) insist on the fact that priors chosen for the fit are the primary drivers, before the type of SFH, to recover the physical parameters. The size of priors thus sets strong constraints on the ability to run fitting codes, especially for large samples. The speed of CIGALE allows it to explore various sets of priors with several $10^8$ models in a reasonable time for thousands of spectrophotometric objects. The periodic SFH is chosen because it is conceptually different from the delayed-plus-burst SFH: it does not assume any kind of continuous SFH. Instead, a series of bursts, separated by regular quiescent periods, is used.

---

[2] The delayed-plus-burst SFH is defined as: $SFR(t) \propto t/\tau^2 \times \exp(-t/\tau)+burst(t)$, with t being the time and the burst being a continuous star formation event happening in the last 1, 5 or 25 Myr (*33*).
[3] The periodic SFH is a series of rectangular star formation events, with the following parameters: δ, the elapsed time between the beginning of each star formation episode, τ, their duration, and the age of the onset of the first star formation episode (i.e. the age of the oldest stars).

We validate the line fluxes measured by CIGALE with those estimated via other methods ([Supplementary Materials, Comparison of the fluxes derived by CIGALE with other papers](#)). The comparison is good down to about 3σ of the background.

In order to help estimating the dust mass for these objects, we make use of deep 450 and 850 µm SCUBA-2 and NOEMA-1.1 mm observations. Both are cross-correlated with JWST's coordinates. The former have a mean depth of $\sigma_{450}$=1.9 and $\sigma_{850}$=0.46 mJy beam$^{-1}$ and the angular resolution is $\theta_{FWHM}$≈8 arcsec at 450 µm, and $\theta_{FWHM}$≈14.5 arcsec at 850 µm. For NOEMA, the rms is $\sigma_{1.1\ mm}$=0.10 mJy beam$^{-1}$, and the beam size is 1."35×0."85. While the sensitivity of these observations is typically lower than what would be required to detect most of our galaxies, the inferred upper limits are useful to put constraints on the total IR luminosities and dust masses. Similarly, the angular resolutions are much larger than JWST's (*[13](#)*). Some of the associations might thus be wrong. However, these sub-mm data rule out any strong far-IR emitters that would be associated with the objects in our sample.

Because the far-IR information for this sample is limited, we cannot directly derive any information on the dust emission SED shape. However, the ALMA-ALPINE sample presents physical properties, e.g. stellar masses ($\log_{10}(M_{star})$∼10), redshifts (4.5<z<6.2) similar to ours (*[39](#)*, *[40](#)*). We thus assume we can make use of the same best model identified in (*[39](#)*, *[40](#)*). Although modified black bodies are not used here, we note that the ALMA-ALPINE model corresponds to a dust temperature $T_{dust}$=54.1±6.7 K, assuming an optically thin modified black-body. Using this dust emission best model, the only needed constraints are on the total IR luminosity and the energy balance, that is that the total IR luminosity comes from the dust obscuration and an energy conservation. The information on the amount of dust attenuation comes from the line ratios, especially Hα/Hβ when available, the UV slope $\beta_{FUV}$, and from any available IR/sub-mm data ([Fig. 2](#)).

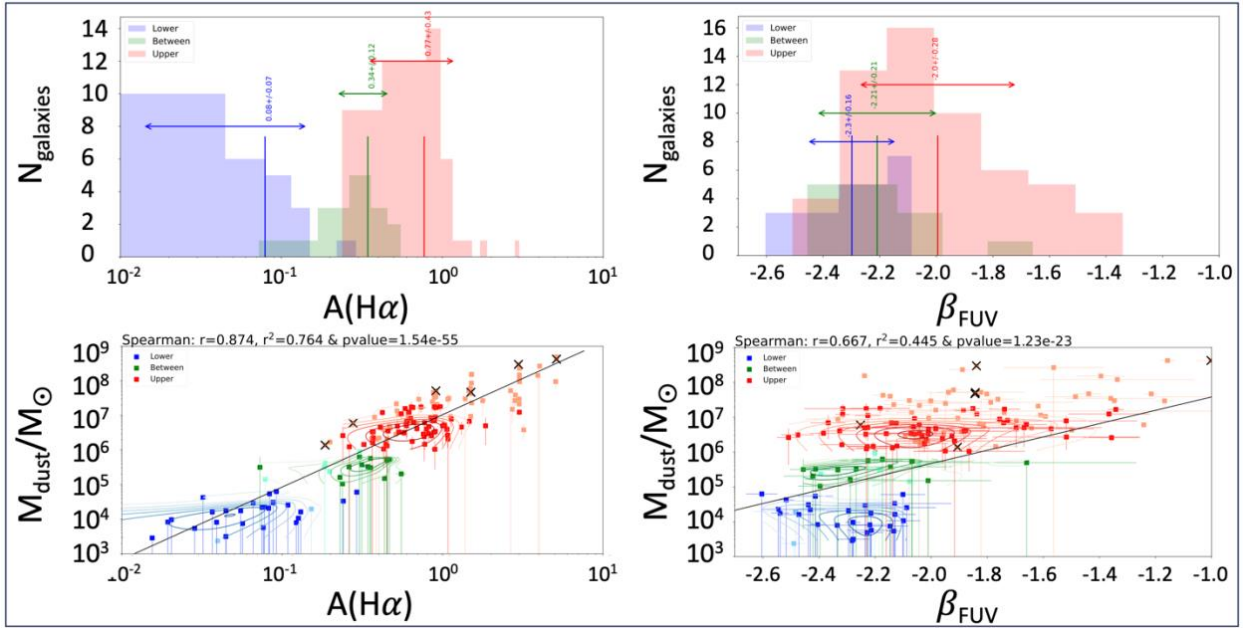

**Fig. 2: Correlation of the dust mass, $M_{dust}$ with the dust attenuation, $A(H\alpha)$ and the UV slope $\beta_{FUV}$.** The most relevant parameter to predict the dust mass is the dust attenuation $A(H\alpha)$ derived from the $H\alpha/H\beta$ Balmer decrement. The correlation with $A(H\alpha)$ alone (lower left) accounts for 76% of the variation in $M_{dust}$, whereas the correlation with $\beta_{FUV}$ alone (lower right) accounts for 44% of this variation. The other tested parameters: metallicity (~ 2%) and redshift (< 1%), but also the level of the sub-mm upper limits are not significantly correlated with $M_{dust}$ in this analysis. The brightest H II regions in local galaxies show a correlation between the Balmer line reddening and the dust mass surface density (47, 48). Our high-redshift galaxies are small (<$RH_{F200W}$>=1.7±0.6 kpc) and probably dense; They might also be dominated by H II regions. The spectral information brought by NIRSpec is fundamental to estimate the dust masses.

## 2. *The coevolution of metals and dust, and the critical metallicity*

Details on how the metallicity, the mass of metals and gas are computed as well as comparisons with other estimates are given in the Supplementary Materials (Technical details on estimating the metallicities, the mass of metals, and the mass of gas).

We show in Fig. 3 the trends[4] related to the rise of the specific mass of metals ($M_Z/M_{star}$) and that of dust ($M_{dust}/M_{star}$) with cosmic age, and with the star formation rate sSFR=SFR/$M_{star}$. Several points can be noticed: first, both follow a similar trend; second, $M_Z/M_{star}$ is always higher than $M_{dust}/M_{star}$: the mass of metals is larger than the dust mass; Third, we observe a lack of extremely low-metallicity galaxies: all the galaxies observed so far are above a critical metallicity value, $Z_{crit}=10^{-6}-10^{-4}$ $Z_\odot$, required for the Pop.III to Pop.II transition, that is from the formation of high-mass to that of low-mass stars (41, 42, 43). This might be due to the scarcity of such pop.III objects that makes them difficult to detect with small-sized surveys, and/or to a very fast rise of metals in the early Universe.

---

[4] We observe the same trends for a periodic SFH (Supplementary Materials, Results assuming a periodic star formation history), suggesting this is not an SFH-dependent result.

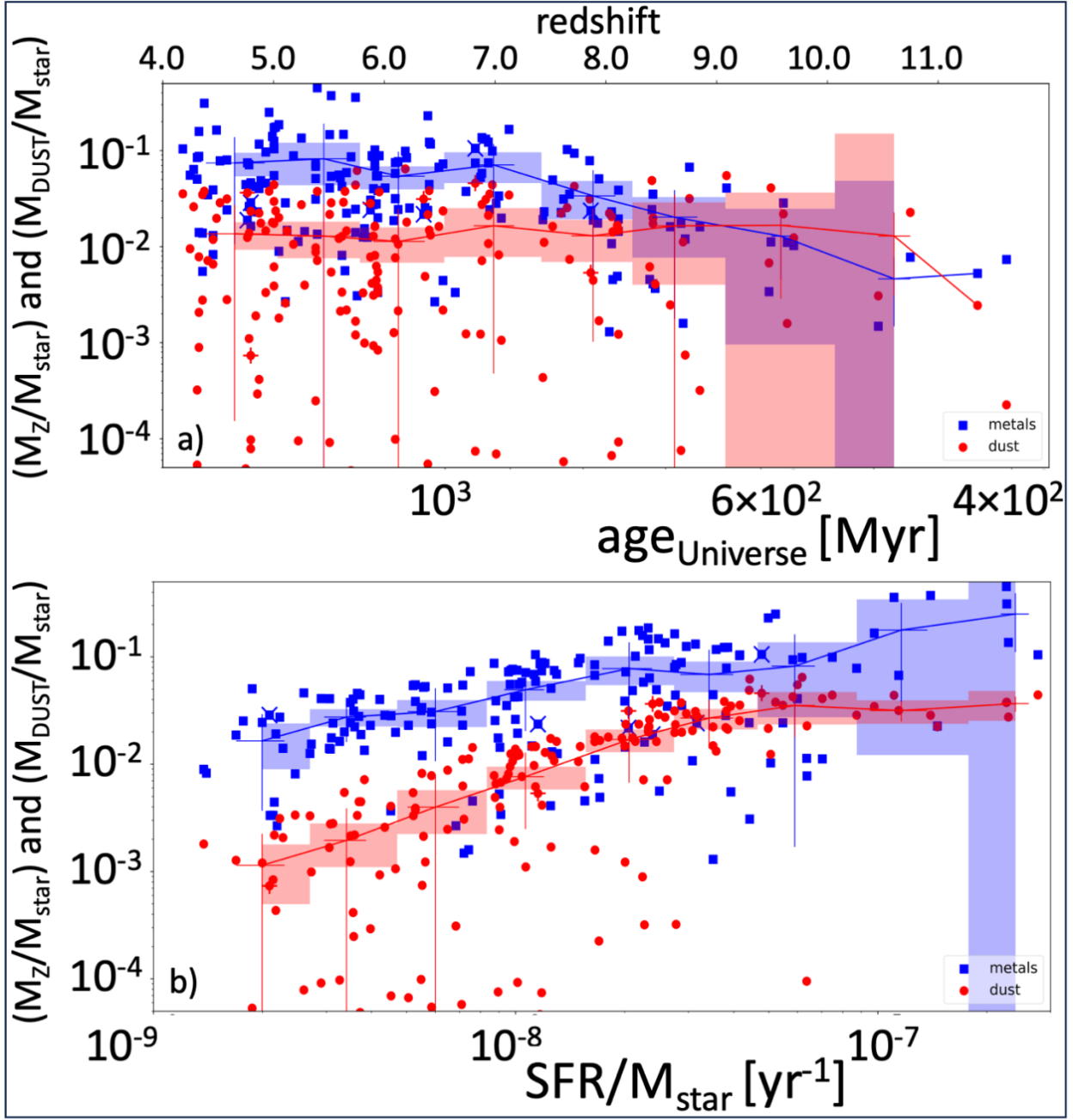

**Fig. 3: The evolution of the mass of metals and mass of dust:** a) $M_Z/M_{star}$ (in blue) and $M_{dust}/M_{star}$ (in red) both regularly increase with the age of the Universe: more metals and dust grains are formed as the Universe ages. We note that even for galaxies in the early Universe (age$_{Universe}$≲600 Myr or z≳9), $M_Z/M_{star}$ never goes below a few $10^{-3}$, possibly suggesting a fast rise of metals that produces the observed threshold. $M_Z/M_{star}$ increases faster than $M_{dust}/M_{star}$, with a much larger dispersion for $M_{dust}/M_{star}$. The light blue/red area show the mean and 2σ confidence interval of the distribution within several bins. b) $M_Z/M_{star}$ (in blue) and $M_{dust}/M_{star}$ (in red) decrease from high to low sSFR (also see [44], [45]) with a larger dispersion, only at lower sSFR for $M_{dust}/M_{star}$.

3.    *The transition from stardust to ISM dust in galaxies*

In Fig. 4, we show the far-UV dust attenuation, $A_{FUV}$, vs. $\log_{10}(M_{star})$ where we again note a large dispersion, similar to that in Fig. 3, at low $M_{star}$ spanning about 2 decades. The ratio of dust-to-stellar mass, $M_{dust}/M_{star}$ as a function of sSFR, color-coded with $A_{FUV}$ (Fig. 4) suggests the large dispersion observed in both figures have the same origin. The stellar mass is not the leading parameter as both low and high $M_{dust}$ lie in the same stellar mass range. This trend appears to be related to the dust mass with a much lower $A_{FUV}$ for $\log_{10}(M_{dust})<5.0$.

Quantifying the mean dust-to-metal, DTM=<$M_{dust}/M_Z$> in Fig. 3, for galaxies with $\log_{10}(M_{dust})\leq 5.0$ (lower), $5.0<\log_{10}(M_{dust})\leq 6.0$ (between), and $\log_{10}(M_{dust})>6.0$ (upper), we find DTM $_{lower}$ = 0.004, DTM $_{between}$ = 0.050, DTM $_{upper}$ = 0.976. There is much less dust in galaxies with low $A_{FUV}$. These objects are those appearing at the bottom left of the $M_{dust}/M_{star}$ vs. sSFR in Fig. 4.

To estimate $M_{dust}$, we use CIGALE's dust emission models (*46*). We reiterate that we do not derive any shape for the IR emission in this paper: because the shape of the IR spectrum is fixed by the ALMA-ALPINE sample at 4.5<z<6.2, and the IR luminosity is estimated assuming the energy balance concept, $M_{dust}$ is constrained by the amount of dust attenuation, and by the main observables that define this dust attenuation data (Fig. 2). More details on how $M_{dust}$ is derived, and what the quality of the estimation process is in (Supplementary Materials: Discussion on the measurement of the dust mass).

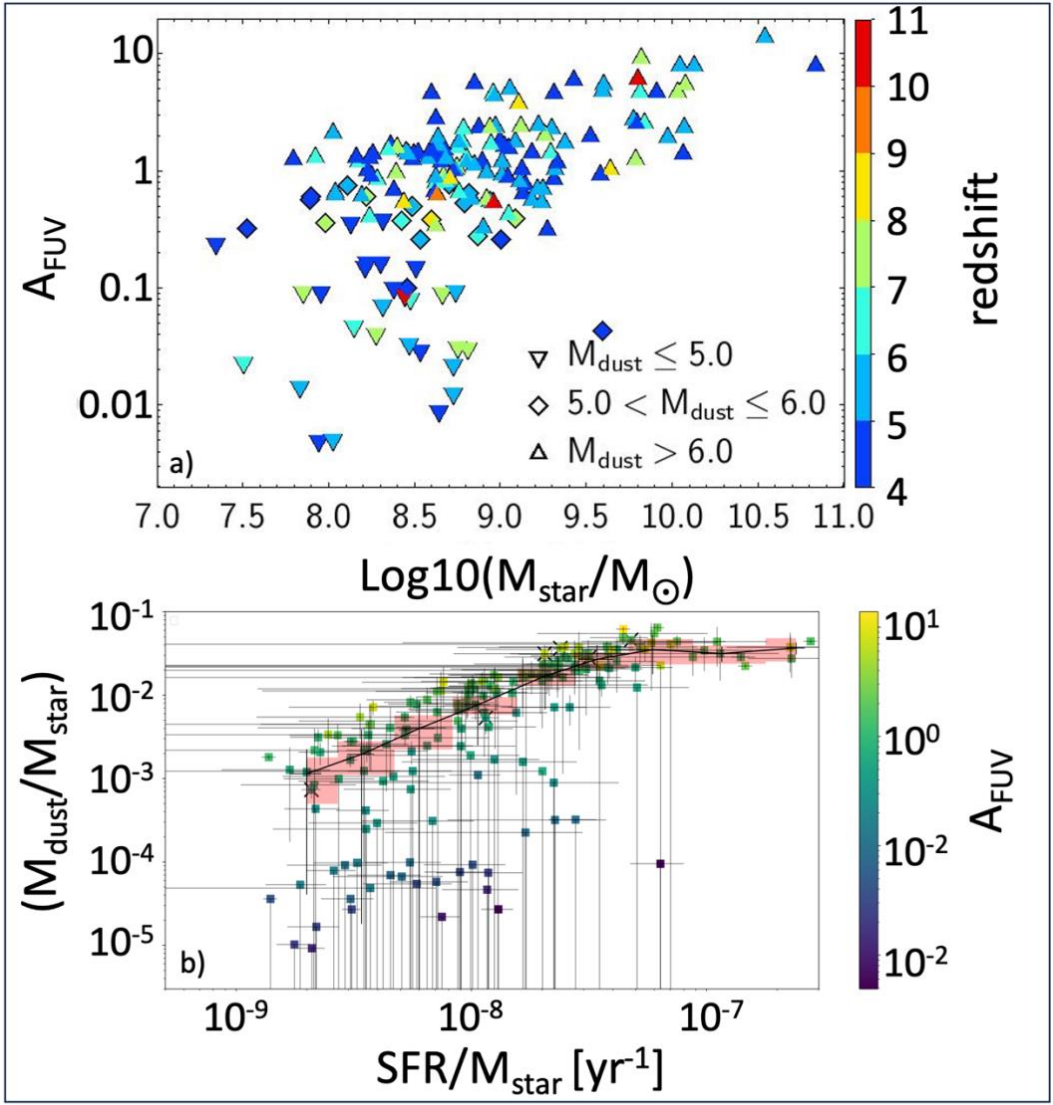

**Fig. 4. Dust properties:** a) $A_{FUV}$ is lower at low stellar mass, and we observe a wide range of $A_{FUV}$ in the stellar mass range $8<\log_{10}(M_{star})<9$, that corresponds to the galaxies with a low dust mass. b) The decrease at sSFR~$10^{-8}$-$10^{-7}$ yr$^1$ is expected (*44*, *45*), and follows the same trend as in Fig. 3, which illustrates the coevolution of metals and dust. At low sSFR, we observe a second sequence that lies about 1-2 dex underneath the primary sequence. In the same sSFR range between $10^{-9}$ and $10^{-8}$ yr$^{-1}$, we both observe attenuated galaxies with a high $M_{dust}/M_{star}$, and galaxies with a very low dust attenuation. The origin of this apparent double sequence will be discussed later. The light-red area shows the mean and 2σ confidence interval of the distribution within several bins.

In the diagram $M_{dust}$ vs. $M_{star}$ plotted in Fig. 5, we observe a tight sequence in the upper part of the figure, mainly in the range $8.0<\log_{10}(M_{star})<10.0$. We also see a significant turn-down at $\log_{10}(M_{star})$~8.0-9.0, that might also be identified in Fig. 4. This never-observed effect means that $M_{dust}$ is significantly lower by a factor of 100-1000 at a given stellar mass, with a lower second sequence, parallel to the upper one. This biphasic plot suggests a two-mode building of dust mass in galaxies.

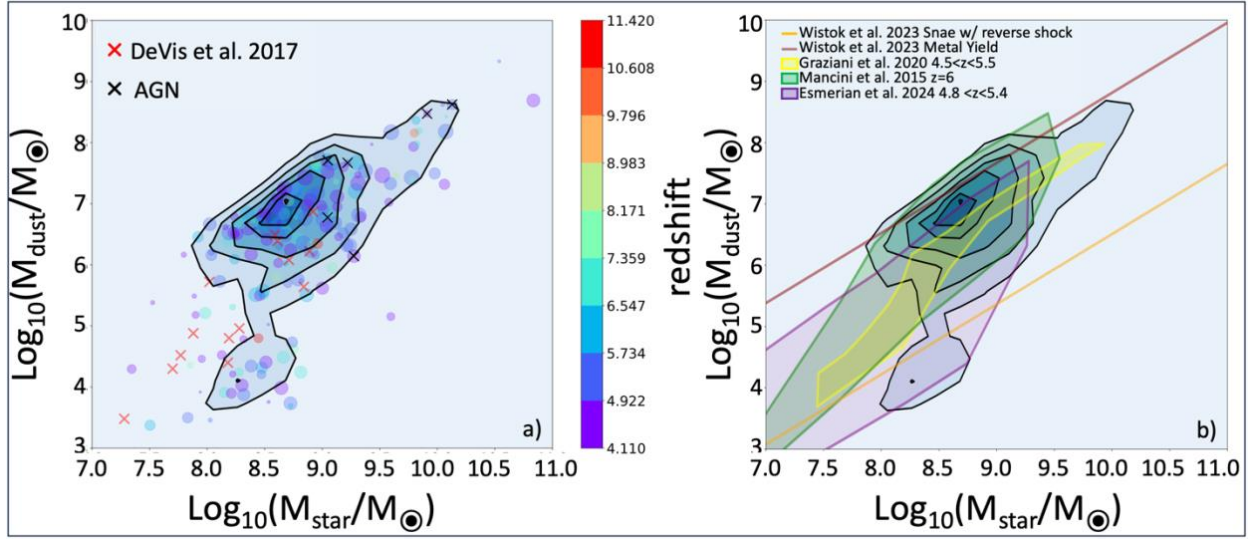

**Fig. 5: $M_{dust}$ as a function of $M_{star}$:** a) At $\log_{10}(M_{star}) \sim 10^{8-9}$ $M_\odot$, we observe, for the first time, a transition from a low to a high sequence. The dots are color-coded in redshift. The size of the symbols provides us with information on $12/\log_{10}(O/H)$ with the largest sizes corresponding to the largest metallicities. We show density contours in black, heavy lines. The red crosses (46) present a subsample of galaxies from an HI-selected sample of local galaxies that might be similar to our low-$M_{dust}$ objects. b) We superimpose models on the density contours. The upper brown line (21, Wistock et al. 2023) shows where galaxies with ISM-grown grains should be. The bottom orange line corresponds to galaxies with only stardust (21, Wistock et al. 2023) that underwent a 95% destruction of grains by reverse SNe shock. The comparison with models: Mancini et al. 2015 (19) in light green, Graziani et al. 2020 (20) in light yellow and Esmerian et al. 2024 (22) in light purple, suggests that the lower sequence corresponds to stellar dust from SNe and from AGB stars. The upper sequence where $M_{dust}$ is larger at a given $M_{star}$ implies that a large part of the dust mass is formed in the ISM.

In order to understand the nature of this lower sequence, Fig. 5 shows several models (19, 20, 21, 21). The first dust grains formed via circumstellar processes linked to stellar evolution in SNe (and maybe AGB stars). However, most of these dust grains are probably destroyed by the SNe reverse shock. After this first phase, the remaining dust grains form seeds, and accrete ISM material for grain growth. This process seems to happen only when a critical ISM metallicity is reached (49). While the upper sequence would have a dust mass where grains have grown in the ISM, the lower sequence would correspond to stardust grains only formed from SNe (and maybe AGB stars), with a grain destruction rate by the SNe reverse shock of the order of 95 %. The jump from the lower to the upper sequence predicted by models perfectly matches what appears in our data (but see 50).

This population of stardust galaxies provides a natural explanation for the excess of UV-bright galaxies at z>10 detected by JWST. If the proportion of stardust galaxies increases when the redshift increases, the dust attenuation would be much lower (for our lower sequence, $A_{FUV} \leq 0.09$, $H\alpha/H\beta = 2.86 \pm 0.07$, and $\beta_{FUV} = -2.31 \pm 0.16$), thus producing more UV light. The present result provides us with an inherent explanation, if galaxies at z>10 only contain a low dust mass mainly formed in the circumstellar medium around SNe in the very first phases of star formation [13]. In our small sample of 111 galaxies in the mass range $7.57 \leq \log10(M_{star}) \leq 8.85$ corresponding to the transition region, the only object at z>10 in the lower sequence is the z=11.4 Maisie's galaxy, suggesting stardust galaxies might be dominant in the early Universe.

## *4. Summary*

We detect a downturn in the $M_{dust}$ vs. $M_{star}$ diagram at $\log_{10}(M_{star}) \sim 8.5$, marking the shift from dust solely produced by stellar evolution (stardust) to dust growth in the ISM of galaxies. This transition aligns with the prediction of dust evolution models.

The galaxies with low $M_{dust}/M_{star}$ and blue UV slopes contain young, metal-poor stars that may be forming their first dust grains from Pop.II—and at z>9, possibly Pop.III—stars, along with their first metals.

Such stardust galaxies would be ideal suspects to produce the excess of UV-bright galaxies in the early Universe [*12*, *13*]. They might be dominant in the early Universe.

We do not detect any extremely low-metallicity values at z>8, suggesting either a bias in our sample, or a rapid rise of metals in the early Universe.

We developed a version of the CIGALE code that fits spectrophotometric data. Tests show it can reliably estimate key emission line fluxes, down to 3σ above the background, and measure gas metallicities.

**Supplementary material**

**Acknowledgments:**

**Funding:**

- DB and VB thank the Centre National d'Etudes Spatial (France) and the Programme National Cosmologie et Galaxies (France).
- VB thanks the Institut Universitaire de France (France).

**Author contributions:**

- Conceptualization: DB
- Investigation: DB, VB, PT, JZ
- Methodology: DB, VB, MD, TD, SF, PAH, DK, PT, JZ
- Software: MB, DB, VF, YR
- Validation: DB, VB, MD, SF, VF, JZ
- Writing – original draft: DB
- Writing – review & editing: DB, VB, PT, MD, SF, VF, EG, BWH, RAL, BM, GR, LMS, JZ
- Visualization: DB, VB, JZ
- CEERS executive committee: AF, HCF, SF, AMK, DK, JSK, JL, CP, NP, RS
- CEERS program architects: MBB, MD, SF, PAH, JSK, AMK, CP, NP
- CEERS key project architects: NC, HCF, AF, AG, NG, PPG, AK, AMK, DK, LK, JL, SR, RS, CT, SW, GY, LYAY

**Supplementary Materials**

Materials and Methods

Supplementary Text

Figs. S1 to S10

Tables S1 to S2

References (*51–69*)

# Science
## AAAS

Supplementary Materials for

**Title: CEERS: Forging the First Dust – Transition from Stellar to ISM Grain Growth in the Early Universe**

Authors: Denis Burgarella[1*], Véronique Buat[1], Patrice Theulé[1], Jorge Zavala[2], Pablo Arrabal Haro[3], Micaela B. Bagley[4], Médéric Boquien[5], Nikko Cleri[6,7,8], Tim Dewachter[1], Mark Dickinson[3], Henry C. Ferguson[9], Vital Fernández[10], Steven L. Finkelstein[11], Adriano Fontana[12], Eric Gawiser[13], Andrea Grazian[14], Norman Grogin[9], Benne W. Holwerda[15], Jeyhan S. Kartaltepe[16], Lisa Kewley[17], Allison Kirkpatrick[18], Dale Kocevski[19], Anton M. Koekemoer[9], Arianna Long[4], Jennifer Lotz[9], Ray A. Lucas[9], Bahram Mobasher[20], Casey Papovich[21,22], Pablo G. Pérez- González[23], Nor Pirzkal[24], Swara Ravindranath[25,26], Giulia Rodighiero[27,28], Yannick Roehlly[1], Caitlin Rose[16], Lise-Marie Seillé[1], Rachel Somerville[29], Steve Wilkins[30,31], Guang Yang[32], L. Y. Aaron Yung[9]

**Affiliations:**

[1]Aix Marseille Université, CNRS, CNES, LAM, Marseille, France

[2]National Astronomical Observatory of Japan, 2-21-1 Osawa, Mitaka, Tokyo 181-8588, Japan

[3]NSF's National Optical-Infrared Astronomy Research Laboratory, 950 N. Cherry Ave., Tucson, AZ 85719, USA

[4]Department of Astronomy, The University of Texas at Austin, Austin, TX, USA

[5]Université Côte d'Azur, Observatoire de la Côte d'Azur, CNRS, Laboratoire Lagrange, 06000, Nice, France

[6]Department of Astronomy and Astrophysics, The Pennsylvania State University, University Park, PA 16802, USA

[7]Institute for Computational and Data Sciences, The Pennsylvania State University, University Park, PA 16802, USA

[8]Institute for Gravitation and the Cosmos, The Pennsylvania State University, University Park, PA 16802, USA

[9]Space Telescope Science Institute, Baltimore, MD, USA

[10]Instituto de Investigación Multidisciplinar en Ciencia y Tecnología, Universidad de La Serena, Raul Bitràn 1305, La Serena 2204000, Chile

[11]Department of Astronomy, The University of Texas at Austin, Austin, TX, USA

[12]INAF - Osservatorio Astronomico di Roma, via di Frascati 33, 00078 Monte Porzio Catone, Italy

[13]Department of Physics and Astronomy, Rutgers, the State University of New Jersey, Piscataway, NJ 08854, USA

[14]INAF--Osservatorio Astronomico di Padova, Vicolo dell'Osservatorio 5, I-35122, Padova, Italy

[15]Physics & Astronomy Department, University of Louisville, 40292 KY, Louisville, USA

[16]Laboratory for Multiwavelength Astrophysics, School of Physics and Astronomy, Rochester Institute of Technology, 84 Lomb Memorial Drive, Rochester, NY 14623, USA

[17]Center for Astrophysics | Harvard & Smithsonian, 60 Garden Street, Cambridge, MA 02138, USA

[18]Department of Physics and Astronomy, University of Kansas, Lawrence, KS 66045, USA

[19]Department of Physics and Astronomy, Colby College, Waterville, ME 04901, USA

[20]Department of Physics and Astronomy, University of California, 900 University Ave, Riverside, CA 92521, USA

[21]Department of Physics and Astronomy, Texas A&M University, College Station, TX, 77843-4242 USA

[22]George P. and Cynthia Woods Mitchell Institute for Fundamental Physics and Astronomy, Texas A&M University, College Station, TX, 77843-4242 USA

[23]Centro de Astrobiologia (CAB), CSIC-INTA, Ctra. de Ajalvir km 4, Torrejon de Ardoz, E-28850, Madrid, Spain

[24]ESA/AURA Space Telescope Science Institute

[25]Astrophysics Science Division, NASA Goddard Space Flight Center, 8800 Greenbelt Road, Greenbelt, MD 20771, USA

[26]Center for Research and Exploration in Space Science and Technology II, Department of Physics, Catholic University of America, 620 Michigan Ave N.E., Washington DC 20064, USA

[27]Department of Physics and Astronomy, Università degli Studi di Padova, Vicolo dell'Osservatorio 3, I-35122, Padova, Italy

[28]INAF - Osservatorio Astronomico di Padova, Vicolo dell'Osservatorio 5, I-35122, Padova

[29]Center for Computational Astrophysics, Flatiron Institute, 162 5th Avenue, New York, NY, 10010, USA

[30]Astronomy Centre, University of Sussex, Falmer, Brighton BN1 9QH, UK

[31]Institute of Space Sciences and Astronomy, University of Malta, Msida MSD 2080, Malta

[32]Nanjing Institute of Astronomical Optics and Technology, Nanjing 210042, China

*Corresponding author. Email: denis.burgarella@lam.fr

**The PDF file includes:**

    Materials and Methods
    Supplementary Text
    Figs. S1 to S10
    Tables S1 to S2

**Materials and Methods**

*Description of the spectro-photometric CIGALE*

The concept of CIGALE[1] was developed in the original paper ([6](#)) where multiwavelength data from the far-UV to the far-IR could be used to derive physical parameters by fitting photometric SEDs. The present public and open version 2022.1, July 4th, 2022 of CIGALE ([33](#)) increases the number of modules (that is, physical processes and models). CIGALE is also one of the fastest SED fitting codes in the world ([51](#)), making it faster than some of the machine-learning-based codes. However, the most important difference of this new version is the possibility to combine spectroscopic data to photometric data (hereafter spectrophotometric data), while conserving CIGALE's ability to fit several hundreds of objects in a reasonable time. For instance, fitting the 173 galaxies using 800 million models from this sample takes about 12 hours on a 48-core computer with 512GB of memory. This means that whatever the parameters derived via the fitting process, these parameters have to be consistent with both photometric and spectroscopic data.

In order to combine the two above data types, we have to normalize the spectrum to the photometry. We provide three options: 1) no normalization: raw data are combined, 2) we integrate the modeled spectra into the filters and estimate a global normalization factor through a $\chi^2$ when the signal-to-noise ratio >5 for the photometric bands used to compute $\chi^2$, and 3) we determine a wavelength-dependent normalization. We stress that normalizing the spectroscopic data to the photometric data could be problematic, if the emission inside the photometric aperture is physically different from the emission inside the spectral slit. In this case the resulting fit might not be realistic because of the different natures of the emitting regions. For instance, a dusty galaxy might present a clear region in the outskirts that could dominate the spectrum, if both observations are not at the same position. Converging would thus be difficult, and a good global fit would not be reached. We thus recommend being careful when combining the spectroscopic and photometric data. For small galaxies like ours, this issue is minimized because we are more likely to observe the same region photometrically, and spectroscopically.

In order to simultaneously fit all the data, we need to inform CIGALE about the (sometimes wavelength dependent) spectral resolution of the spectrometer to create resolution elements corresponding to the instrumental spectral resolution where the models are integrated. This phase is transparent to the users, and is performed during the configuration of the CIGALE environment. A typical CIGALE spectrophotometric run appears similar to a photometric run from the user's point of view.

CIGALE learns that some spectroscopic data has to be taken into account from the configuration file, pcigale.ini, where the following flag should be set to 'True':
> Is there any spectroscopic data to analyze? This spectroscopic dataset will be used in conjunction with any photometric data, and/or equivalents widths, line fluxes or other properties.The answer must be: True or False

---

[1] https://cigale.lam.fr/

```
            use_spectro = True
```

Moreover, the input table must contain the following information:

- "id" that contains an alphanumeric identifier (a different one for each object to be fitted)
- "redshift" that contains the redshift of the objects or "NaN" if redshifts have to be estimated.
- "spectrum" which contains the path to the spectrum
- "mode" that contains the type of spectrum, e.g. "prism" if JWST/NIRSpec data is used
- "norm" where one of the three normalizations should be provided for each object: "none", "global" or "wave". Note that the normalization could be different for each object. If no photometric data is given to CIGALE, "none" should be used, and only the spectrum will be fitted.

**AGN in the studied galaxy sample**

We list in tab. S1 the AGN candidates identified in our sample from the literature.

**Sample of spectral fits**

We present in fig. S1 (for the upper sequence with high $M_{dust}$) and in fig. S2 (for the lower sequence with low $M_{dust}$) a sub-sample of galaxies out of our 173 objects. For each of the galaxies, the entire fit to the sub-mm and a zoom on the fit of the spectrum only is also presented.

*Parameters used in CIGALE's final fit*

All the spectral models computed by this new version of CIGALE[2] (momentarily dubbed as CIGALES) using the selected modules (each one corresponding to a physical emission, tab. S2) are convolved with NIRSpec's prism spectral resolution matched to the observed spectra. This spectral data is added to the photometry to form a modeled spectro-photometric SED. The rest of the process follows the usual flow of CIGALE as described in (*33*). The nebular models have been computed with CLOUDY, as described in (*52*). In this analysis, we normalize the prism spectroscopic data by computing a wavelength-dependent normalization which is computed from photometry (NIRCAm and NIRSpec here) in the wavelength range, for the photometric bands that have a SNRε5.0. A second-order polynomial is used to derive the wavelength-dependent normalization factor. This normalization is computed for each and every object with photometric data, and applied to each spectroscopic observation. In this work, we used the WMAP7 cosmology (*53*). We assumed a Chabrier initial mass function (IMF, *54*), and a solar metallicity $Z_\odot = 0.014$ (*55*), and the dust emission models, uses $\kappa_v = 0.637$ m$^2$ kg$^{-2}$.

---

[2] https://gitlab.lam.fr/cigale/cigale.git (clone with HTTPS).

**Supplementary Text**

*Comparison of the fluxes derived by CIGALE with other papers*

We check in fig. S3 and fig. S4 that the flux of the emission lines measured by CIGALE are consistent with first, a more traditional line fitting[3], and second, a fit performed on the sub-sample where both prism and grating NIRSpec spectra are available.

*Technical details on estimating the metallicities, the mass of metals and mass of gas*

  1.   *Methodology and equations*

In CIGALE, the stellar $Z_{star}$ and nebular $Z_{gas}$ metallicities have different priors, and are separately constrained by fitting the spectra, including emission lines. For this specific estimation process, we make use of the new nebular models described in (*52*) that can have excitation parameters up to logU=-1.0, and with a wide range of nebular metallicities, and electronic densities. The metallicity $Z_{gas}$ derived by CIGALE can be converted into $12+\log_{10}(O/H)$, (where O/H is the oxygen abundance of the gas) by using their table 1 (correspondence between $\zeta_0$, the interstellar gas metallicities, and the stellar metallicities). $\zeta_0$ is defined on the oxygen abundance: $\zeta_0 = (O/H) / (O/H)_{GC}$, where $(O/H)_{GC} = 5.76 \times 10^{-4}$, and GC is the so-called local Galactic concordance (*52* follow *58*). From this, Eq. 1 links the total metallicity to the oxygen abundance:

*Eq. 1: $12+\log_{10}(O/H)) = \log10(Z_{gas}) + 10.410$.*

Our CIGALE-derived metallicities are compared to those estimated in other papers (Supplementary Materials., Comparison of the metallicities derived by CIGALE with other estimates)

Fig. 2 presents the specific mass of metals ($M_Z/M_{star}$) for the present sample as a function of the specific SFR (SFR/$M_{star}$), where the metal mass, $M_Z$ is computed from Eq. 2 (*59*):

*Eq. 2: $M_Z = M_{gas} \cdot 10^{12+\log_{10}(O/H)-8.69} \cdot Z_\odot$*

In Eq. 2, the gas mass, $M_{gas}$, and the oxygen abundance, $12+\log_{10}(O/H)$, are estimated via the spectrophotometric fitting and from Eq. 1[4]. For $M_{gas}$, we need to estimate the molecular mass $M_{molgas}$ from Eq. 3 (*60*). The specific instantaneous SFR, sSFR=SFR$_{inst}$/M$_{star}$, is derived from the spectrophotometric fitting, while the reference sSFR for the main sequence (MS) as a function of the stellar mass $M_{star}$, and the redshift: sSFR(MS, z, $M_{star}$) is from (*61*), assuming the so-called "Bluer" w/ high-z obs" MS in their Table 9 (Eq. 4). The MS is modeled up to z~6 (*61*), while our sample reaches z=11.4. However, even though the scatter in the MS is quite large, studies suggest that it should not show any strong evolution to z~12 (*62*, *63*).

---

[3] https://ceers-data.streamlit.app/

[4] Note that the constants in the equation from (*47*) includes a correction of the H$_2$ masses upward by $\alpha_{heavy}$=1.36 for the content of helium and heavy elements, to get a census of the entire mass content of the molecular phase.

***Eq 3:*** *$log_{10}(M_{molgas}) = 0.06 - 3.33 \cdot [log_{10}(1+z) - 0.65]^2 + 0.51 \cdot log_{10}(sSFR/sSFR(MS, z, M_{star},) - 0.41 \cdot [log_{10}(M_{star}) - 10.7] \cdot M_{star}$*

***Eq. 4:*** *$log_{10} SFR(M_{star}, Age_{Universe}) = [(0.73 - 0.027 \cdot Age_{Universe}) \cdot log_{10}(M_{star}) - (5.42 + 0.42 \cdot Age_{Universe})] - log_{10}(M_{star})$*

To estimate $M_{gas}$ we need to add the contribution from the atomic gas $M_{atomgas}$ to $M_{molgas}$ The atomic-to-molecular mass ratio $M_{atomgas}/M_{molgas}$ is estimated by (*64*) for star-forming galaxies at z~0, z~1.0, and z~1.3 with values in the range 4±2 for galaxies with $M_{star}>10^{10}$ $M_\odot$. To account for galaxies with $<10^{10} M_\odot$ in their statistics, they assume that the ratio $M_{atomgas}/M_{molgas}$ are systematically higher by a factor of ~5 at low $M_{star}$. In this case, the values obtained would increase $M_{molgas}$ at z~1.3 by a factor of ~2, yielding a ratio $M_{atomgas}/M_{molgas}$=2.5 for the highest redshifts. At larger redshifts (0.01<z<6.4), there is no significant redshift evolution of the $M_{atomgas}/M_{molgas}$ ratio (*65*), which is about 1-3. At z=8.496, the gas and stellar contents of a metal-poor galaxy is studied with JWST and ALMA (*66*). From this analysis, they infer $M_{molgas}=(3.0-5.0) \cdot 10^8 M_\odot$. corresponding to 40%±10% of $M_{gas}$ for their object, which leads to $M_{atomgas}/M_{molgas}=1.5^{2.0}_{1.3}$. Given the redshift of our objects, we will assume in this paper $M_{atomgas}/M_{molgas}=2.0.^{2.5}_{1.3}$.

## 2. *Comparison of the metallicities derived by CIGALE with other estimates*

In [fig. S5](#) and [fig. S6](#), we compare our CIGALE metallicities estimated for the same CEERS galaxies in (*67*, *68*, *69*). Nakajima et al. (*68*) measured emission-line fluxes for 10 galaxies with [O III] λ4363 Å lines, and determined their electron temperatures, in a way similar to lower-redshift star-forming galaxies. From this, they derive the metallicities of the 10 galaxies by a direct method and also of other JWST-observed galaxies with strong lines using their previous metallicity calibration (*67*) based on the direct-method measurements. Sanders et al. (*69*) also combine JWST measurements with [O III] λ4363 Å auroral-line detections from JWST/NIRSpec and from ground-based spectroscopy to derive electron temperature ($T_e$) and direct-method oxygen abundances on a combined sample of 46 star-forming galaxies at z=1.4-8.7. Even though we use a nebular emission model designed to reproduce known scaling of emission line ratios with ISM properties including metallicity, our fitting method is considered different, as the total spectro-photometric fits allow to consistently constrain the metallicities by selecting only models that are in agreement with the whole information brought by observations (continuum and lines, together).

### *Results assuming a periodic star formation history*

We present the same analysis obtained in the main paper, but here, we assume a periodic SFH, that is a series of regular bursts over the age of the galaxies. The main parameters that define the SFH for this CIGALE run are listed in [tab. S2](#). The conclusions presented in the main paper could also be reached with the periodic SFH, confirming that the type of SFH does not fundamentally impact the results of the paper.

*Discussion on the measurement of the dust mass*

From the best-fit parameters, CIGALE can create a mock catalog based on the best models. To do so, CIGALE uses the best models for each of the galaxies in our sample, built from the best-fit parameters. The observed noise is randomly added to the best-fit spectrophotometric data. Then, we refit these simulated data with the exact same set of priors used for the initial fit. Note that 10% of additional random error based on the observed uncertainties is also added to the simulated data. When comparing the input "exact" data to the output Bayesian data, we are able to estimate whether or not the output derived physical parameters can be trusted, and what are the limits. [Fig. S9](#) shows that we could recover dust masses down to about $\log_{10}(M_{dust}/M_\odot)\sim5.0$.

The analysis of the results suggests that the SCUBA-2 sub-mm fluxes do not significantly help in constraining the dust mass, as we do not find any correlation between the measured fluxes or upper limits. The NOEMA detections provide flux densities that are more useful to constrain the dust mass. However, we only have two objects in the sample, and none of them within the lower sequence.

As already discussed in [Sect. 3](#), the Balmer decrement H$\alpha$/H$\beta$ and the dust attenuation for H$\alpha$, A(H$\alpha$) are strongly correlated with $M_{dust}$ (correlation coefficient $r_{H\alpha/H\beta}=0.874$), and the UV slope $\beta_{FUV}$ is also, although at a lower level, correlated with $M_{dust}$ (correlation coefficient $r_{\beta FUV}=0.667$). We can thus conclude that first, the emission lines, and second, the continuum shape drive the estimation of the amount of energy transferred into the far-IR. For our galaxy sample, the observed spectrum seems to bring the main information to estimate this IR luminosity, and $M_{dust}$, if the IR spectrum is assumed to be known.

**Figures**

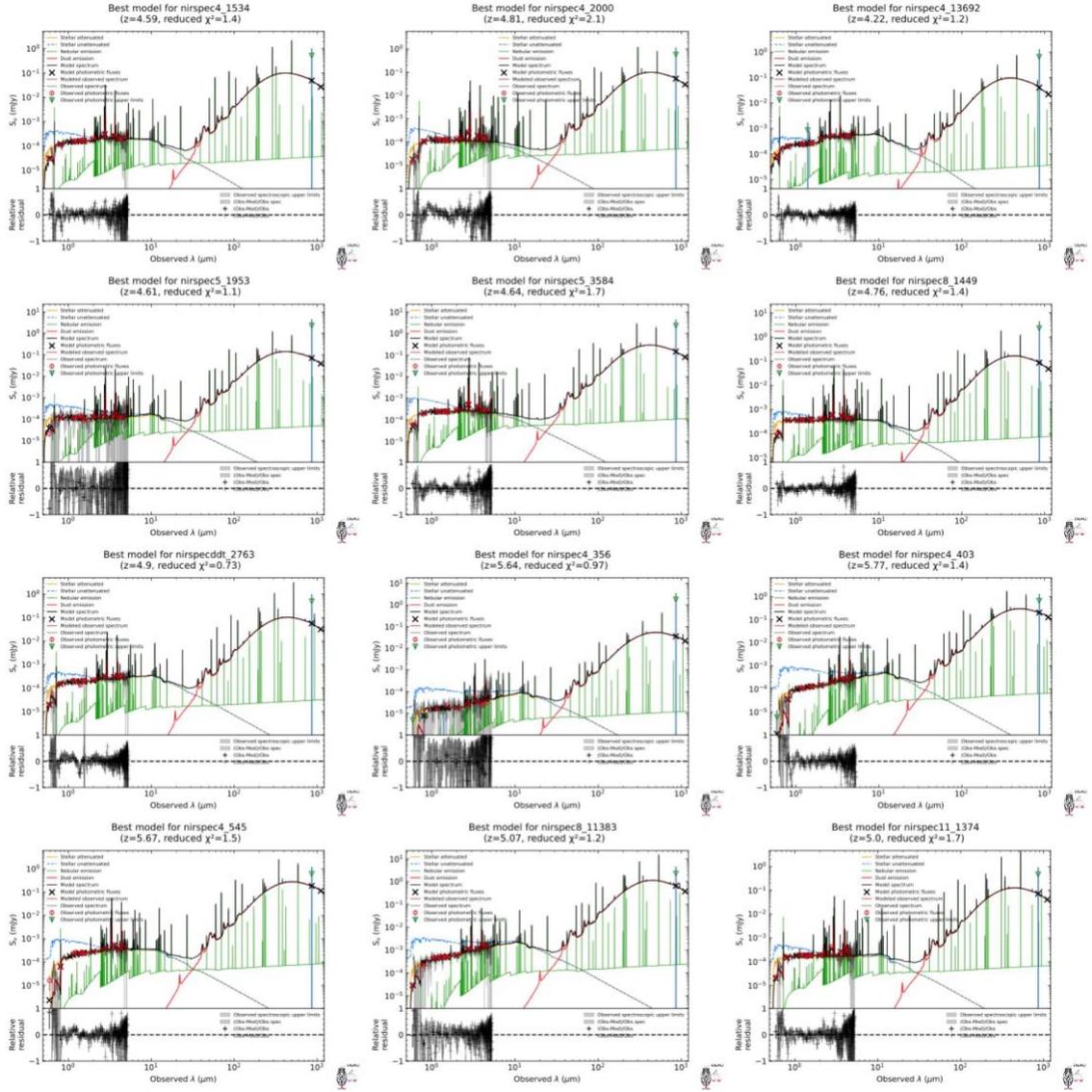

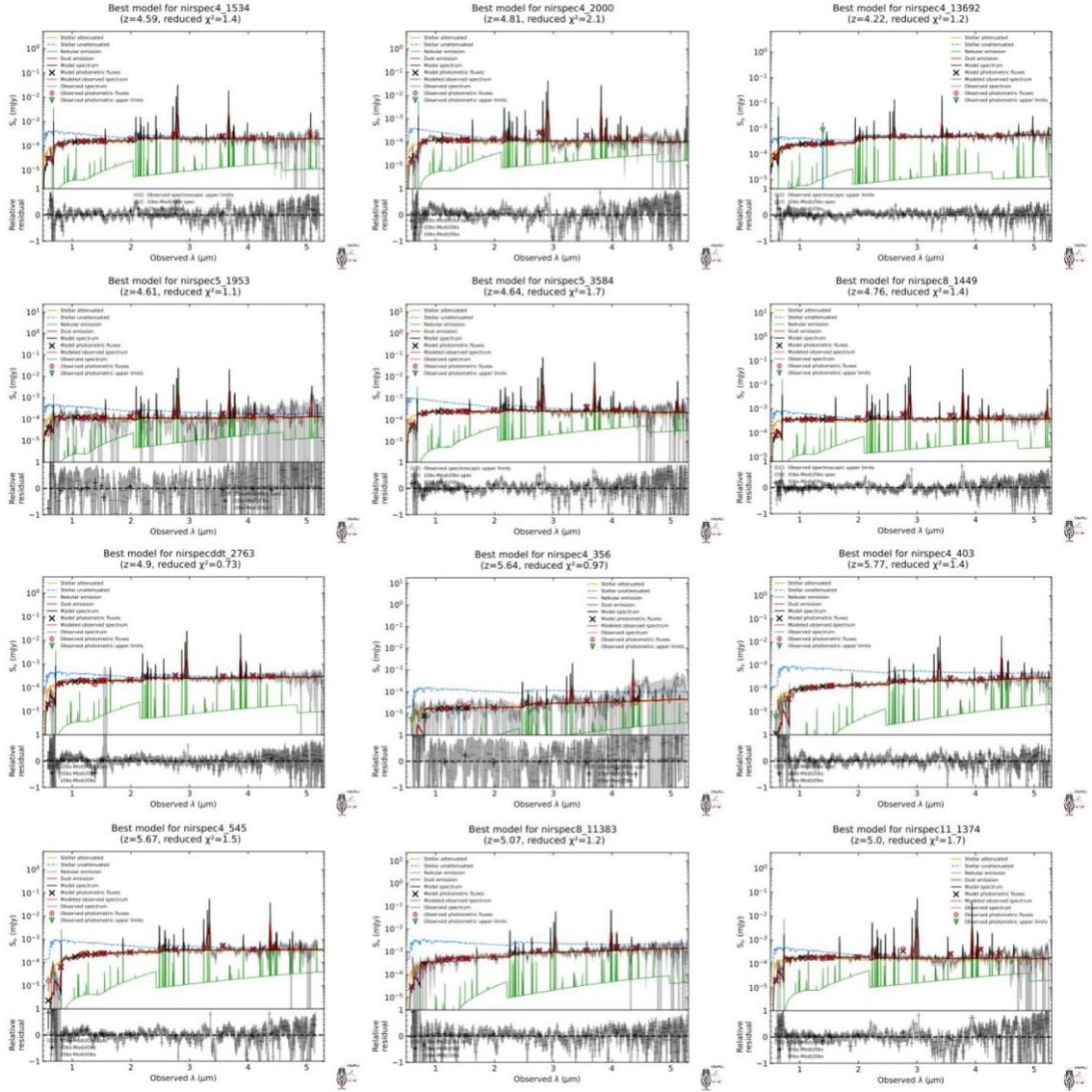

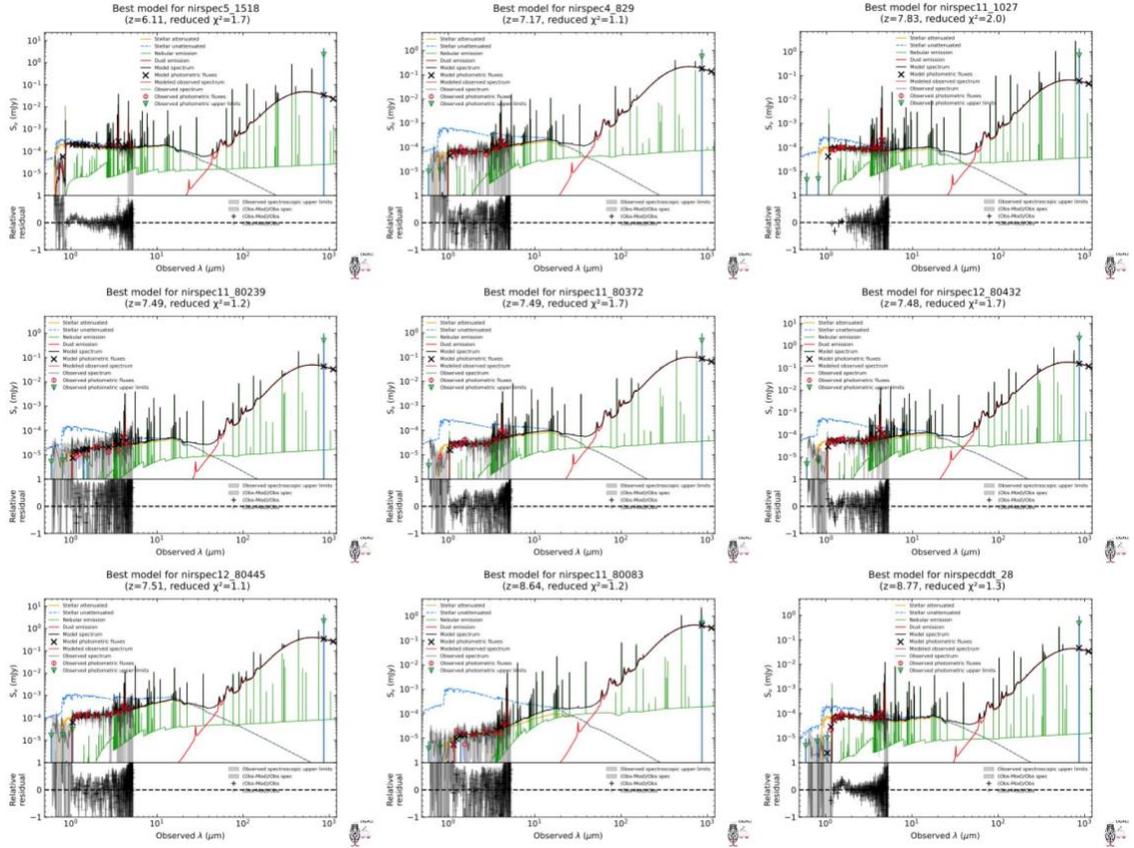

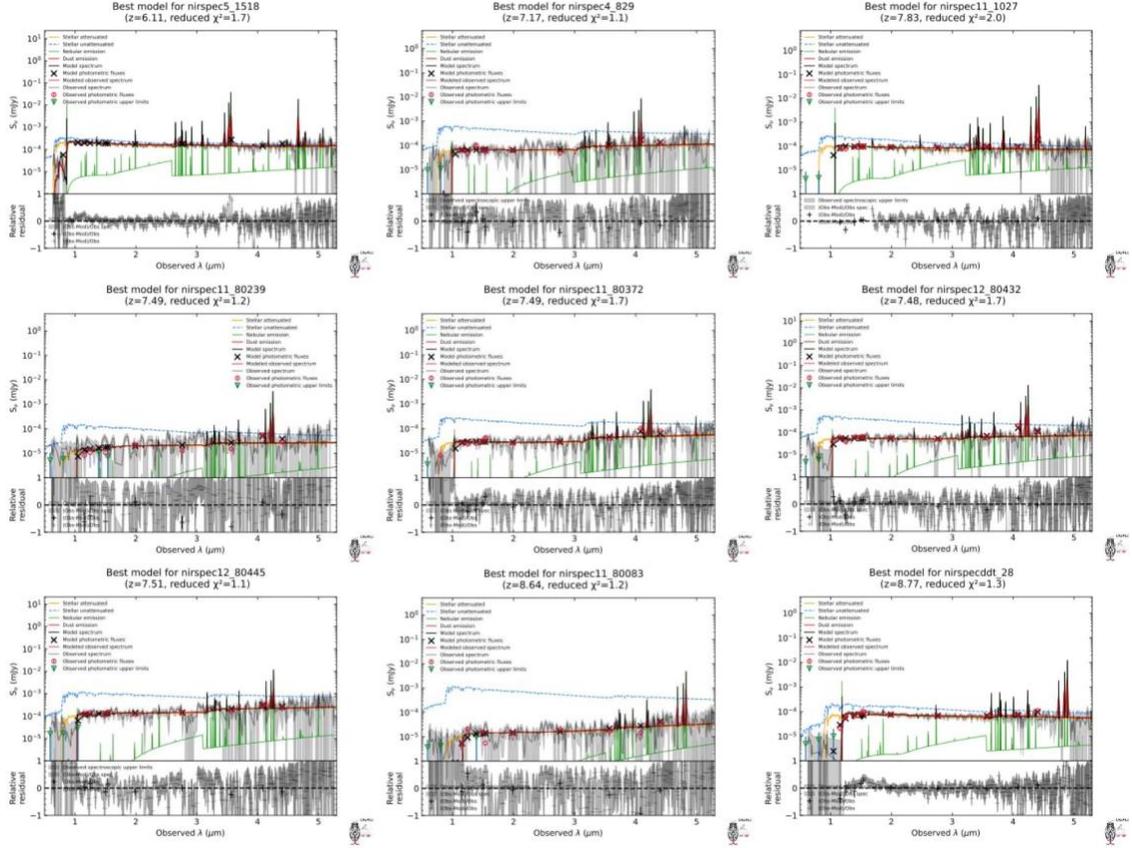

**Figure S1: Objects in the upper sequence (larger $M_{dust}$) in $M_{dust}$ vs. $M_{star}$ plot.** We present a sample of spectral fits over the whole spectral range, that is including NIRSpec spectroscopy and the sub-millimeter data (mostly upper limits). We also show for the same objects the fits of the NIRSpec spectrum alone, which is a zoom in the previous plots.

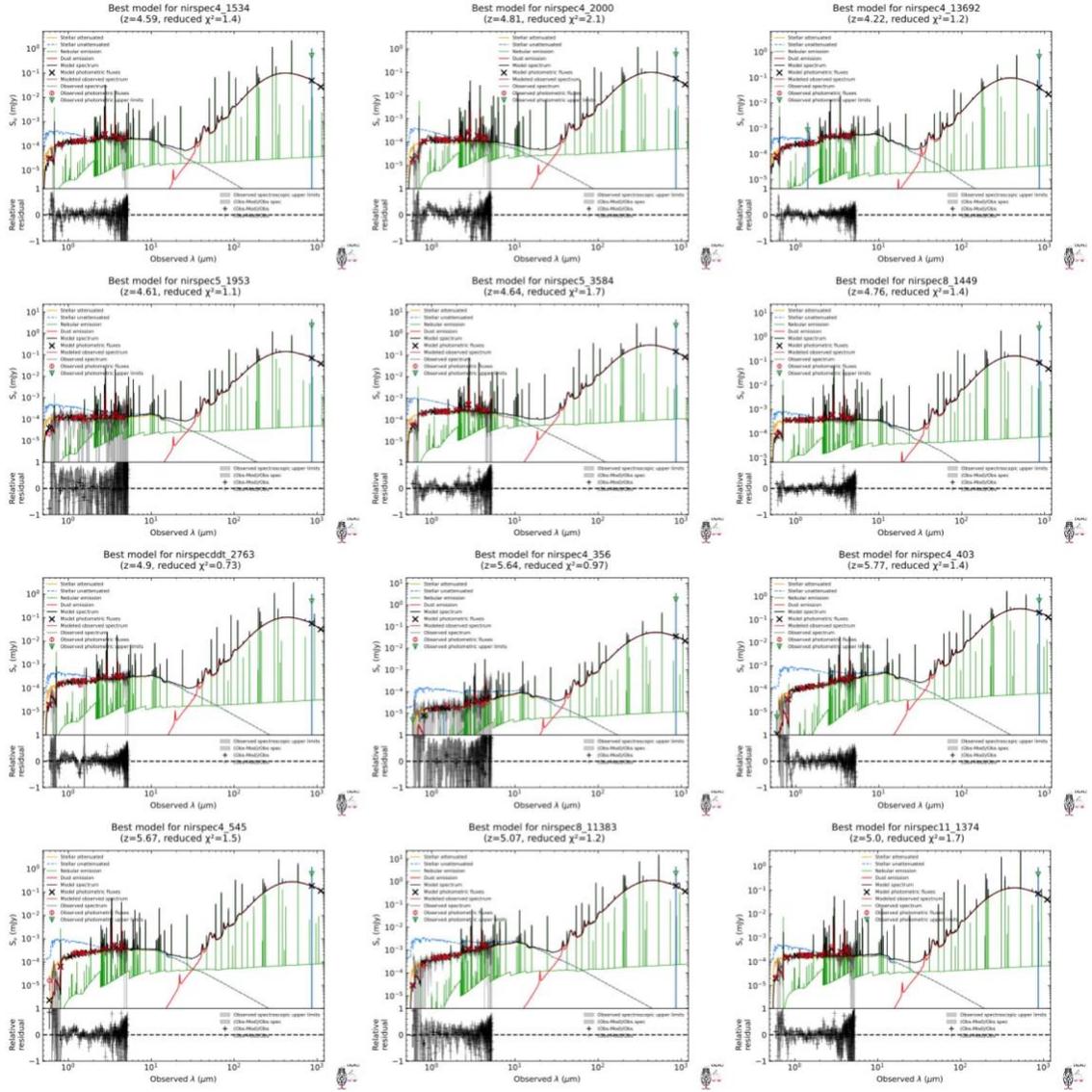

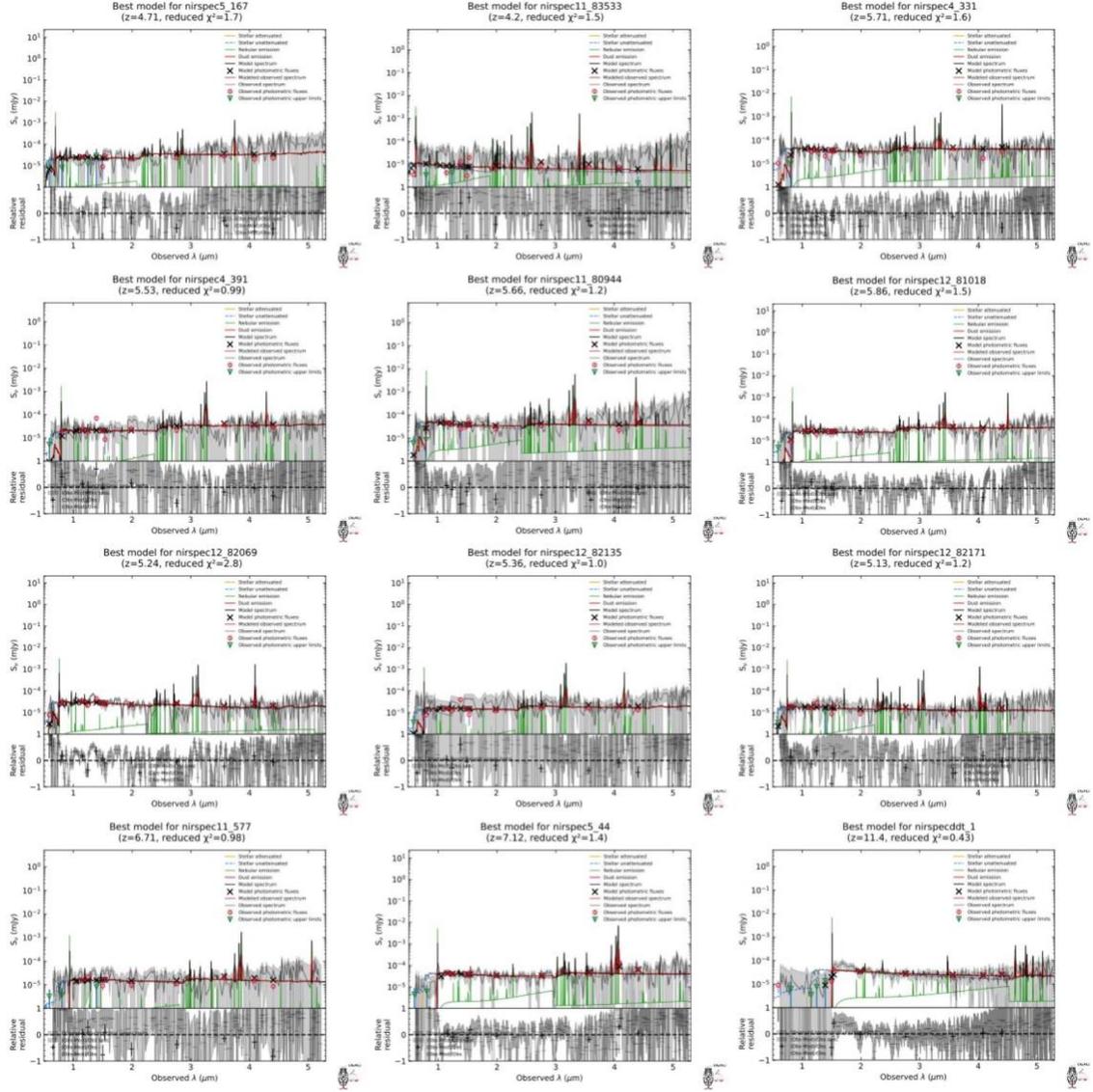

**Figure S2: Objects in the upper sequence (lower $M_{dust}$) in $M_{dust}$ vs. $M_{star}$ plot.** Same as figure S1 but for objects in the lower sequence in the $M_{dust}$ vs. $M_{star}$ plot.

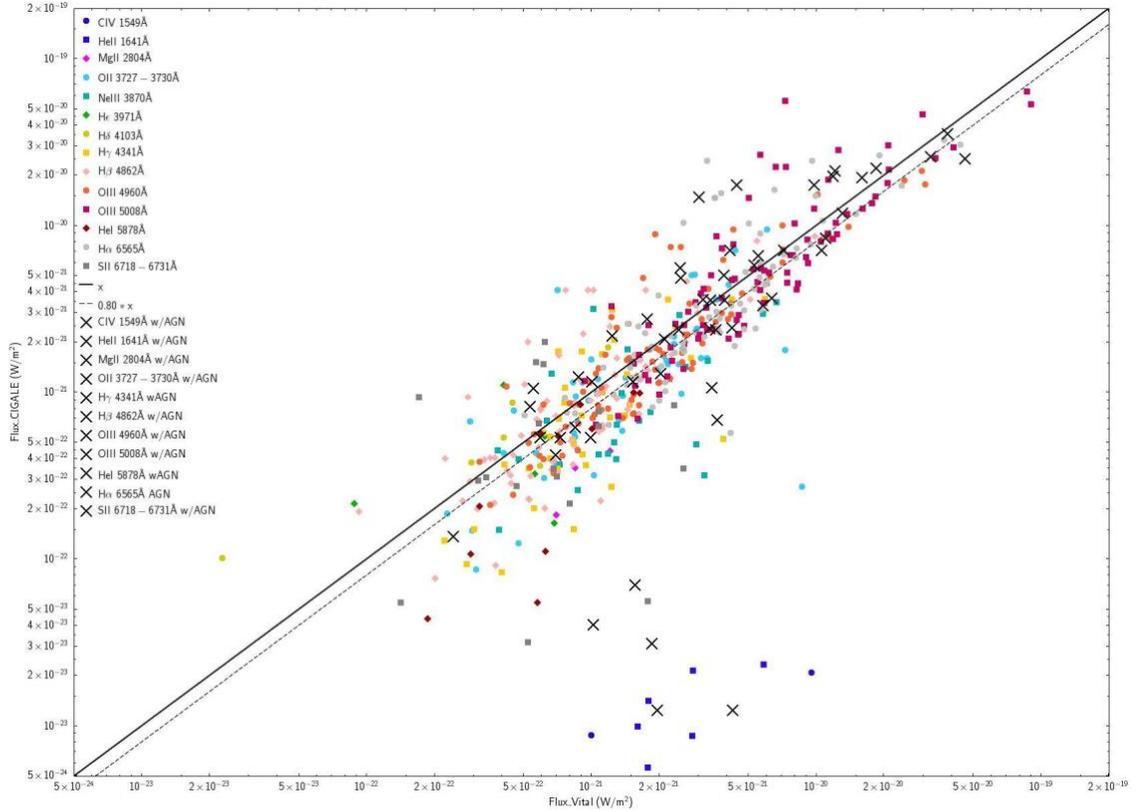

**Figure S3: Comparison with the Gaussian fluxes measured with LiMe** (see Fernández et al. 2024, 57). The full results on the CEERS catalog redshifts and line measurements will be discussed in Arrabal Haro et al. (in preparation). The points corresponding to each emission line are color-coded to better identify each species. The most ultraviolet lines (CIV λ1549 Å, HeII λ1641 Å) in the lower part of the plot, are not in agreement with those estimated (57). The other ones are systematic under-estimated by CIGALE by about 25 %.

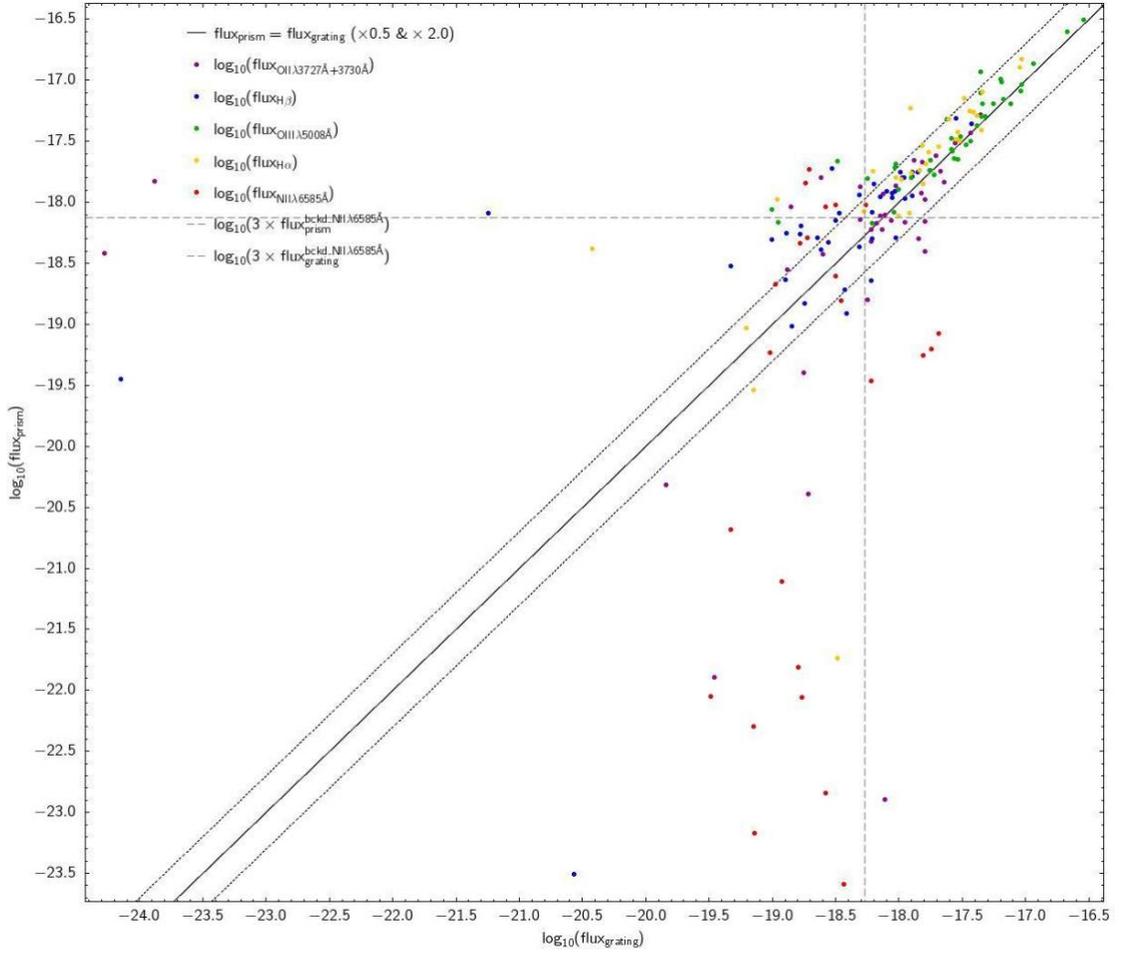

**Figure S4: Comparison of the fluxes computed using CIGALE with Gaussian fitting of the grating** for the objects observed in both configurations. The horizontal and vertical dashed lines show the level of the 3σ background around the [NII]+Hα lines.

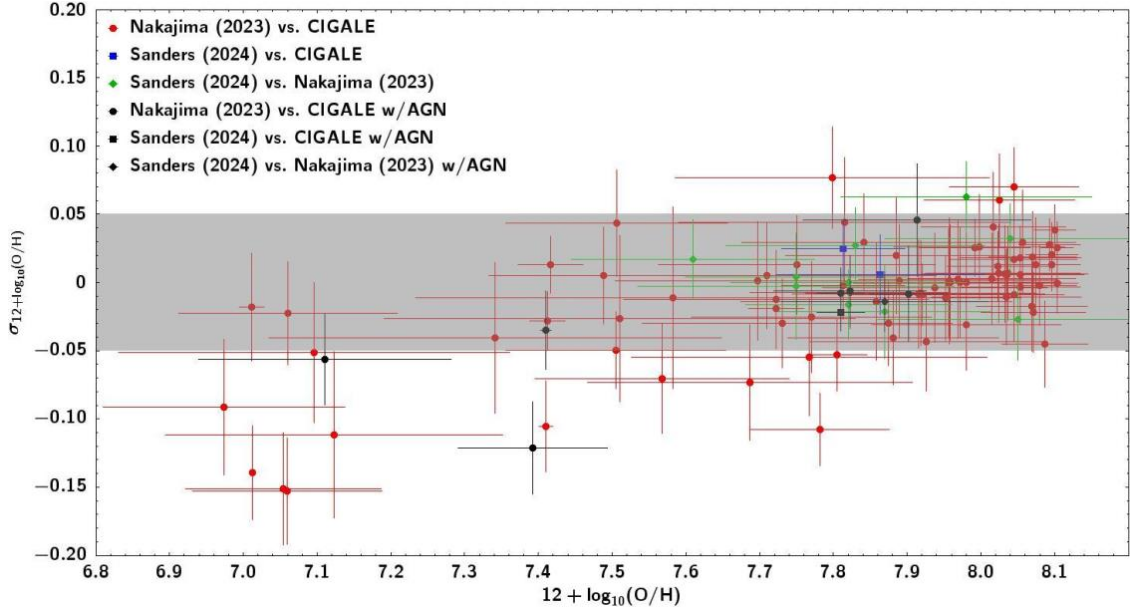

**Figure S5: Difference between metallicity estimates:** Comparison of the metallicities measured for some of our galaxies (*67*, *68*, *69*) via auroral lines with our own estimates. At relatively high metallicities, above 12+log$_{10}$(O/H)~7.4-7.6 (that is Z/Z$_{gas}$~0.1-11% Z$_\odot$), the differences remain within σ(12+log$_{10}$(OH))<0.05. This is about the same dispersion between other metallicity estimates (*67*, *68*, *69*), although on a smaller sample. However, there is a disagreement at lower metallicities, and CIGALE's 12+log$_{10}$(OH) presents an offset that might be systematic or increasing with lower metallicities by about σ (12+log$_{10}$(OH))<0.10. But the direct method is not well calibrated at 12+log$_{10}$(OH)<7.4-7.6 because there are less than 5 objects with auroral lines at such low metallicities estimated via the direct-method oxygen abundances.

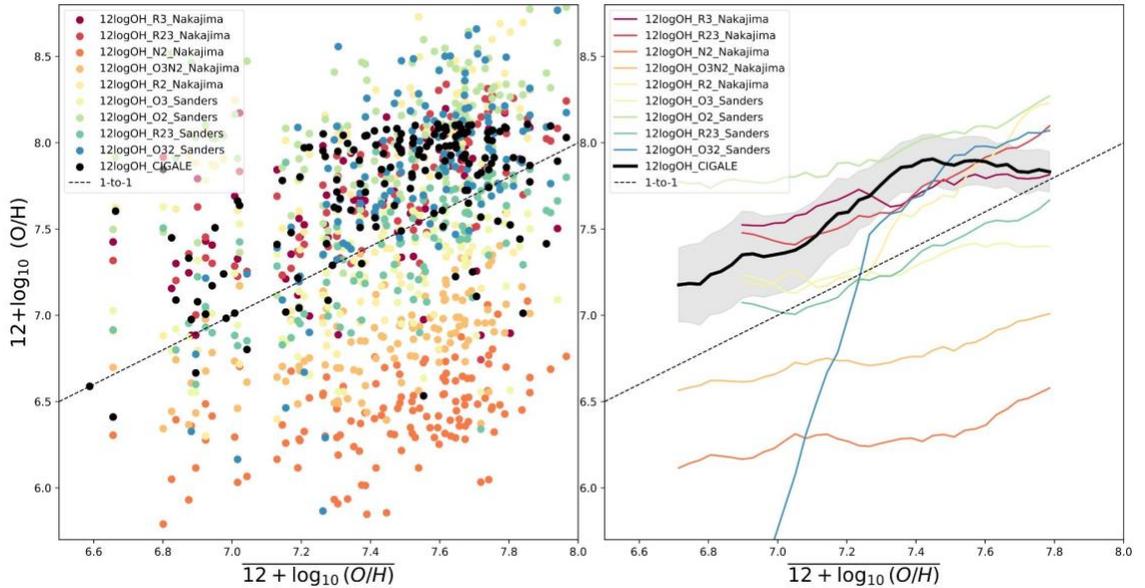

**Figure S6: Comparison of metallicities:** The left panel shows the metallicities derived using various published calibrations as a function of the mean metallicities computed from all the estimates, as listed in the legend. The black one is the output from CIGALE. The dashed black line is the 1-to-1 relation. We can see that for a given metallicity on an x-axis, we could get a very wide range on the y-axis with $\sigma(12+\log_{10}(OH))\sim 2.0$. The right panel shows the same information with a rolling average that shows that the values of $12+\log_{10}(OH)$ estimated by CIGALE are in good agreement with some methods (*67*, *68*) but quite different from the ones from others (*69*). The advantage of CIGALE is that all the spectrophotometric information is combined to provide us with some kind of "summarized value".

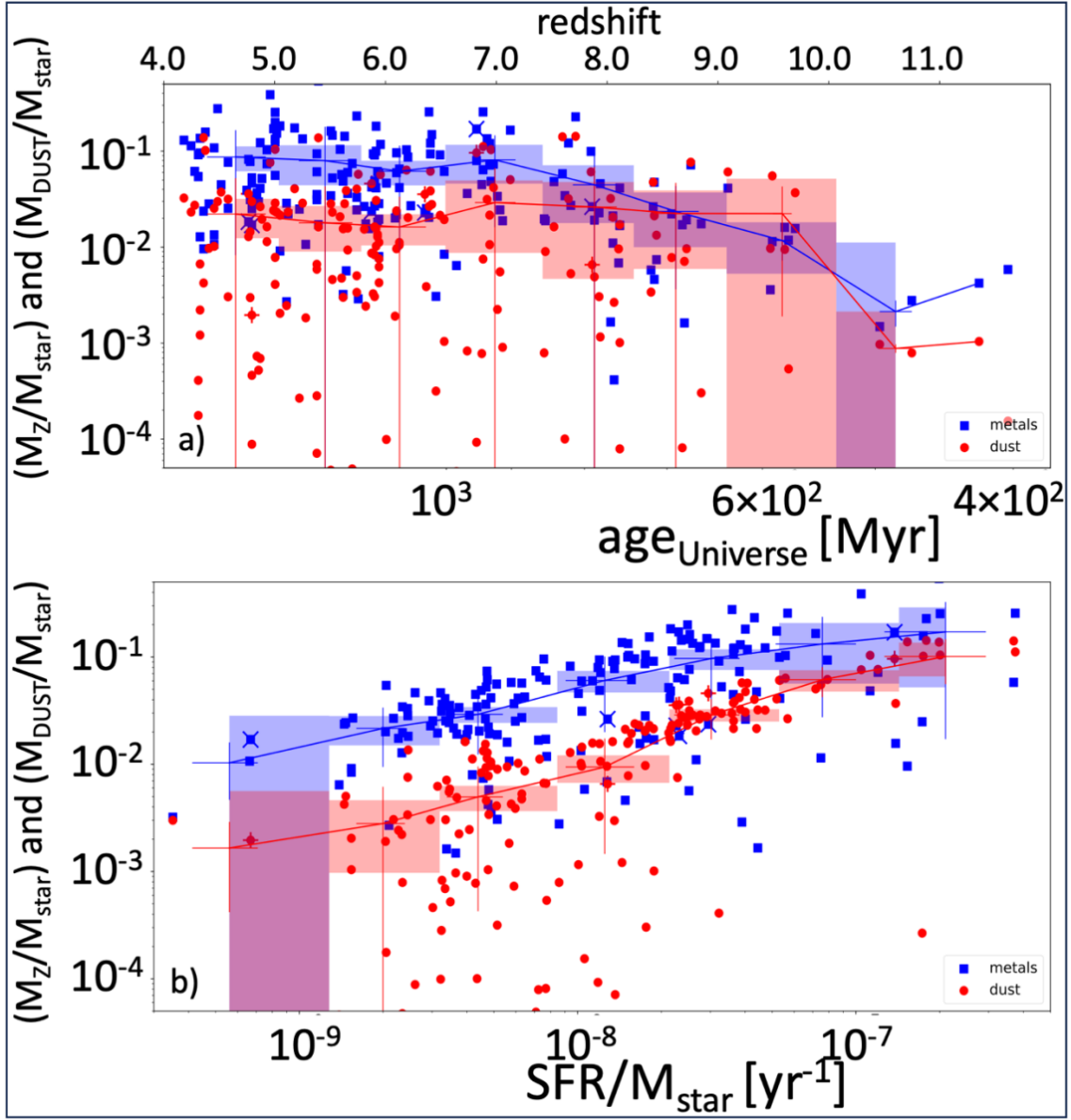

**Figure S7: evolution of the metal-to-stellar mass ratio and dust-to-stellar mass ratio:** this figure is similar to Fig. 2 but with a periodic SFH instead of a delayed plus burst SFH.

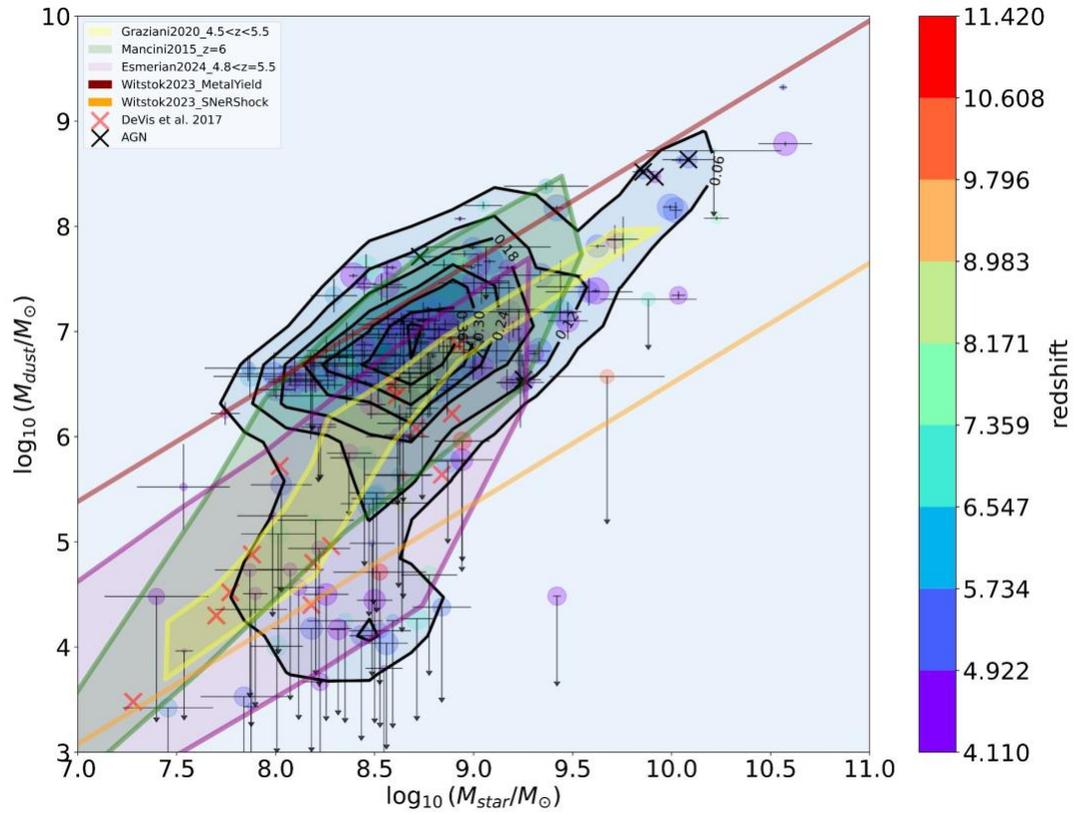

**Figure S8: Test on the stability of the results with a periodic SFH.** It shows two sequences and is very similar to Fig. 3, in the main paper.

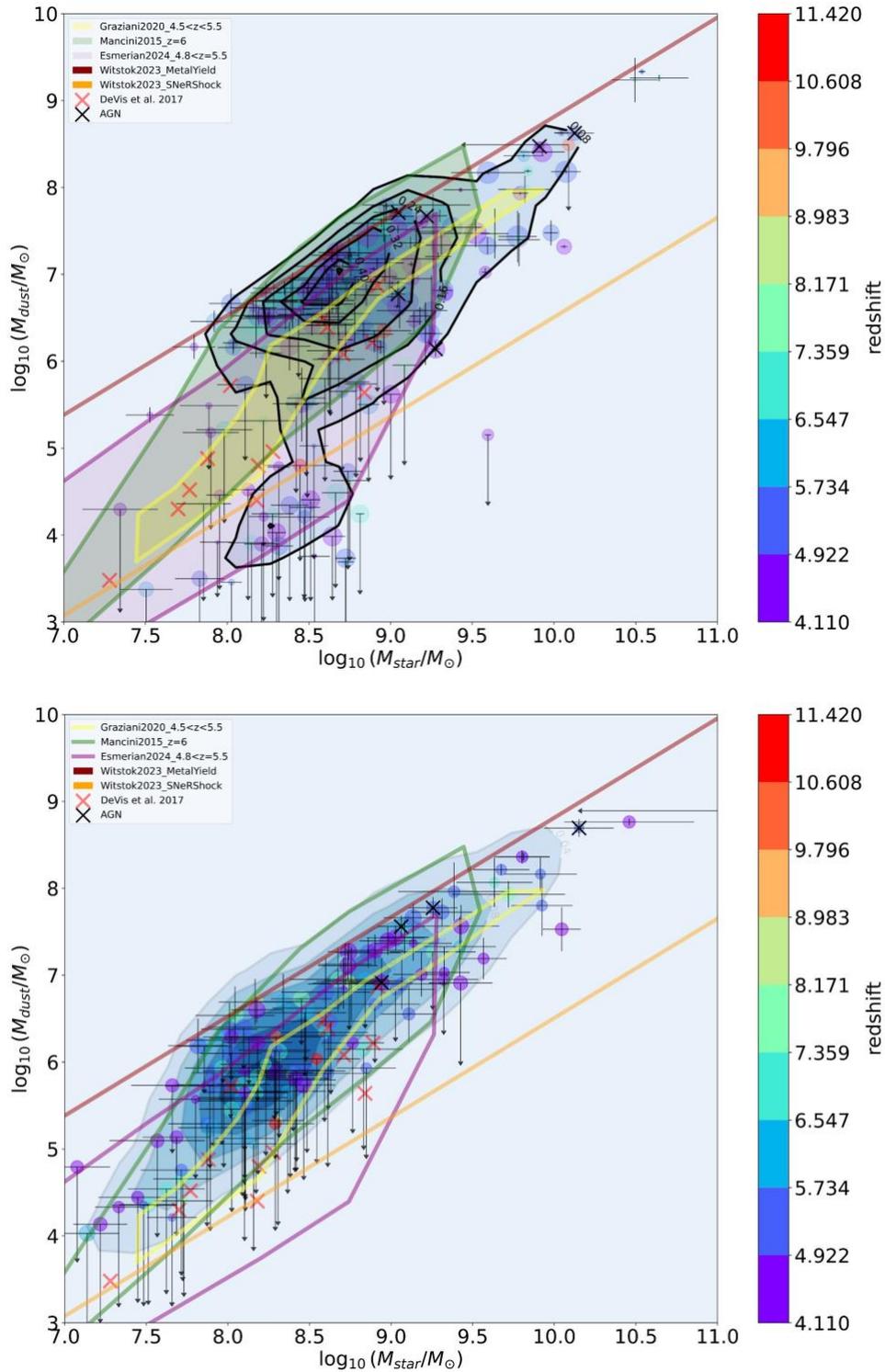

**Figure S9: Test on the stability of the results with the data used:** Top - This figure is created by fitting the spectrophotometric data, as in Fig. 3 in the main paper, except that we do not use the submm ones. The trend observed in this case is almost identical, which confirms the less important role of the sub-mm data in separating the two parallel

sequences. Bottom - This figure is created by only fitting the photometric data, including the sub-mm ones but not the spectroscopic one. The trend observed in this case is different from when we make use of the NIRSpec spectra. We still do see a small decrease at lower stellar mass, even though the less-marked downturn suggests that without spectroscopy, some strong spectral information, and especially the line ratios, is missing. However, the photometric data still bring an information on the dust attenuation because of the UV slope $\beta_{FUV}$. The correlation of $\beta_{FUV}$ with the dust mass is much less significant, and leads to this smaller difference in dust mass, even at low stellar masses which makes the second lower sequence less prominent.

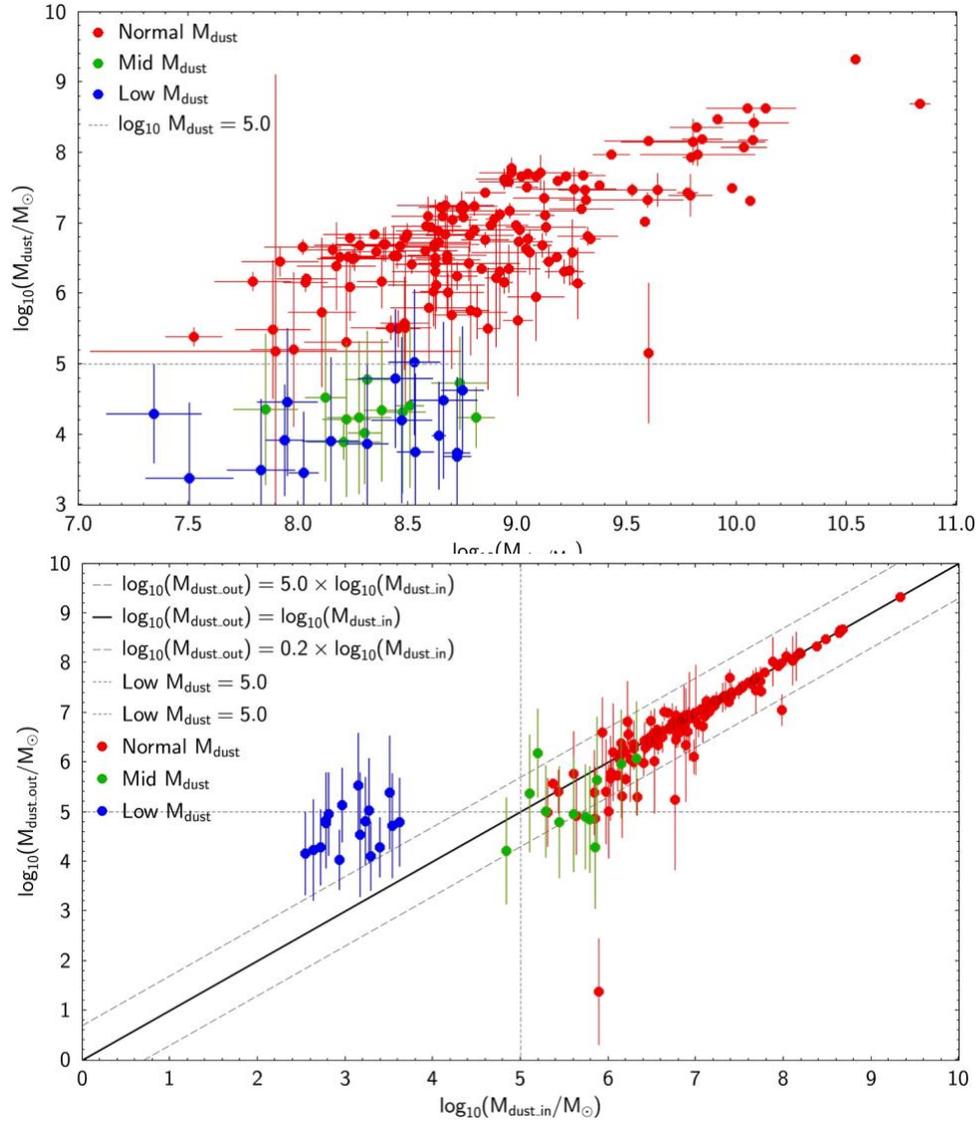

**Figure S10: Use of a mock analysis to estimate the minimum estimated dust mass:** The x-axis shows the modeled input parameters computed from the best-fit models and for each of our galaxies. CIGALE is able to recover (y-axis) by fitting the mock data, these input dust masses, down to about $\log_{10}(M_{dust})\sim 5.0$. Parts of the objects in green and all the objects in blue should thus be considered as upper limits. These objects that are the ones identified as transitioners between the upper sequence and a possible lower sequence, are well below the main sequence (red dots).

**Tables**

| CEERS id | redshift | R.A. | Dec. |
|---|---|---|---|
| nirspec4_397 | 6.01 | 14:19:20.69 | +52:52:57.7 |
| nirspec8_717 | 6.94 | 14:20:19.54 | +52:58:19.9 |
| nirspec4_746 | 5.63 | 14:19:14.19 | +52:52:06.5 |
| nirspec8_1236 | 4.50 | 14:20:34.87 | +52:58:02.2 |
| nirspec7_1244 | 4.48 | 14:20:57.76 | +53:02:09.8 |
| nirspec4_2782 | 5.26 | 14:19:17.63 | +52:49:49.0 |

**Table S1: List of AGN in the analyzed sample.**

| Parameter | Symbol | Run #1 | Run #2 |
|---|---|---|---|
| Sample | CEERS | 173 NIRSpec galaxies | |
| **Star Formation History (SFH)** | | | |
| Type of SFH | | Delayed plus burst | Periodic |
| e-folding time of main stellar population | $\tau_{main}$ [Myr] | 500 | 10, 100 |
| Age of main stellar population | $Age_{main}$ [Myr] | 50, 100, 200, 400, 600, 800, 1000 | 1, 5, 10, 50, 100, 200, 400, 600, 800, 1000 |
| e-folding time of burst | $\tau_{burst}$ [Myr] | 10000 | — |
| Age of late burst | Age [Myr] | 1, 5, 25 | — |
| Mass fraction of late burst population | $f_{burst}$ | 0.0, 0.1, 0.2 | — |
| Elapsed time between the beginning of each event | $\Delta t$ [Myr] | — | 10, 50, 100, 250 |
| **SSP (BC03)** | | | |

| Initial Mass Function | IMF | Chabrier |
|---|---|---|
| Metallicity | $Z_{star}$ | 0.0001, 0.0004, 0.004, 0.008, 0.02, 0.05 |

## Nebular emission

| Ionization parameter | $\log_{10}U$ | -1.0, -1.3, -1.6, -1.9, -2.2, -2.5, -2.8, -3.1, -3.4, -3.7, -4.0 |
|---|---|---|
| Gas metallicity | $Z_{gas}$ | 0.0001, 0.0004, 0.001, 0.002, 0.0025, 0.003, 0.004, 0.005 |
| Electronic density | $n_e$ | 10, 100, 1000 |

## Dust attenuation law

| E_BV_lines, the color excess of the nebular lines light for both the young and old population | E_BV_lines | 1e-7, 1e-4, 0.001, 0.010, 0.050, 0.10, 0.15, 0.20, 0.30, 0.50, 0.75, 1.0, 1.5, 2.0, 5.0 |
|---|---|---|
| Reduction factor to apply on E_BV_lines to compute E(B-V)s, the stellar continuum attenuation. | E_BV_factor | 0.44 |
| Slope delta of the power law modifying the attenuation curve | powerlaw_slope | -0.60, -0.30, 0.00, 0.30 |
| Extinction law to use for attenuating the emission lines flux. | Ext_law_emission_lines | SMC |

## Dust Emission (DL14)

| Mass fraction of PAH | qpah | 0.47 |
|---|---|---|
| Minimum radiation field | Umin | 17 |
| Power Law slope $dU/dM \propto U^\alpha$ | $\alpha$ | 2.4 |
| Fraction illuminated from $U_{min}$ to $U_{max}$ | $\gamma$ | 0.54 |

## No AGN module

**Table. S2: CIGALE modules and input parameters** used for all the fits. BC03 means Bruzual & Charlot (2003, 56) and the Chabrier IMF refers to Chabrier (2003, 54).